\documentclass[prd,twocolumn,superscriptaddress,nofootinbib,longbibliography,floatfix]{revtex4-2}
\usepackage{graphicx}
\usepackage{amsmath,amssymb}
\usepackage{booktabs}
\usepackage{placeins}
\usepackage{xcolor}
\usepackage{url}
\usepackage[T1]{fontenc}
\usepackage[utf8]{inputenc}
\usepackage{microtype}
\usepackage[colorlinks=true,linkcolor=blue,citecolor=blue,urlcolor=blue]{hyperref}
\graphicspath{{./}{figures/}{../results/}{results/}}
\newcommand{\sv}{\langle\sigma v\rangle}
\newcommand{\dl}{\Delta(-2\ln\mathcal{L})}
\newcommand{\code}[1]{\texttt{#1}}
\newcommand{\fermi}{\textit{Fermi}}

\newcommand{\orcid}[1]{\begingroup\hypersetup{hidelinks}%
\href{https://orcid.org/#1}{\IfFileExists{orcid.pdf}{\includegraphics[width=10pt]{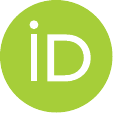}}{\textsuperscript{\scriptsize iD}}}\endgroup}
\begin{document}
\title{Foreground dependence of the 20\,GeV Galactic halo gamma-ray excess and its proposed connection to the LUX-ZEPLIN event}
\author{Kevork N. Abazajian\orcid{0000-0001-9919-6362}}
\email{kevork@uci.edu}
\affiliation{Department of Physics and Astronomy, University of California, Irvine, California 92697, USA}
\author{Youyou Li\orcid{0000-0002-5152-2971}}
\email{y.li4@uva.nl}
\affiliation{GRAPPA Institute, University of Amsterdam, 1098 XH Amsterdam, The Netherlands}
\author{Oscar Macias\orcid{0000-0001-8867-2693}}
\email{macias@sfsu.edu}
\affiliation{Department of Physics and Astronomy, San Francisco State University, San Francisco, CA 94132, USA}
\affiliation{GRAPPA Institute, University of Amsterdam, 1098 XH Amsterdam, The Netherlands}

\date{\today}
\begin{abstract}
We assess the dark matter interpretation of the extended gamma-ray excess near 20\,GeV reported by Totani and its proposed connection to the LUX-ZEPLIN (LZ) high-energy nuclear-recoil event, which independently favors dark matter masses above approximately 200\,GeV.
Using fifteen years of \fermi-LAT data, we reproduce the excess under the original foreground assumptions and spatial binning, and then test it against 281 Galactic diffuse emission configurations, more than forty times that in the original analysis, together with three descriptions of the \fermi\ bubbles, four of Loop~I, and uniform and X-ray-brightness-weighted templates of the eROSITA bubbles.
A component correlated with a Navarro--Frenk--White (NFW) annihilation template remains when the spatial binning is refined and the instrumental response is included. 
Outside the \fermi\ bubbles footprint, the NFW-correlated component is detected with nominal significances of approximately $4$--$16\sigma$ over 5.6--130\,GeV, depending on the adopted foreground model, and of approximately $1$--$7\sigma$ when the enlarged bubble boundaries of a recent substructured \fermi\ bubbles model are also excluded. 
However, both the amplitude and spectrum of the excess depend strongly on the foreground model, and neither the structured nor the flat \fermi\ bubbles template yields a single NFW normalization across the sky. 
Allowing additional spatial freedom in the eROSITA bubbles can reduce the apparent north/south asymmetry, but the excess still cannot be described by a single NFW normalization across different regions of the sky. 
With the substructured \fermi\ bubbles model, which improves the fit substantially, an NFW-correlated component remains at $7$--$18\sigma$ in four of six representative foreground configurations but not in the other two, and the partition rotated by $45^\circ$ still disfavors a single NFW normalization in each configuration tested. 
Simulations also show that mismodeling of the Galactic foreground can produce spurious NFW-correlated emission with an amplitude comparable to that observed, and that constructing the \fermi\ bubbles template from the fit residuals can bias the recovered spectrum. 
The particle interpretation is similarly foreground dependent. 
In particular, the 420\,GeV $W^+W^-$ spectrum motivated by the proposed connection to the LZ event provides an acceptable description for some foreground models but is disfavored by others. 
Taken together, these results disfavor interpreting the full gamma-ray excess as dark matter annihilation in a smooth NFW halo, and weaken the present case for a common dark matter origin with the LZ event.
However, they do not exclude a dark matter contribution to the persistent halo-like residual, and whether such a contribution connects to the LZ event depends on a better understanding of the large Galactic structures with which it is entangled.
\end{abstract}
\maketitle

\section{Introduction}
\label{sec:intro}

Gamma rays from the Milky Way provide a powerful probe of particle dark
matter. If dark matter annihilates into Standard Model particles, it can
produce an extended Galactic signal whose spectrum depends on the particle
physics and whose morphology traces the dark matter distribution
\cite{Slatyer:2017sev,Fermi-LAT:2012pls,Chang:2018bpt}. In practice,
however, these searches are limited by the complexity of the Galactic
gamma-ray sky. Emission from cosmic-ray interactions with interstellar gas
and radiation fields overlaps the expected dark matter signal, as do
unresolved sources and large-scale structures such as the Fermi bubbles.
For this reason, a significant preference for a halo template is not, by
itself, sufficient to establish a dark matter origin. The inferred signal
must also remain reasonably stable as the Galactic foreground model is
varied.

Totani \cite{Totani:2025fxx} recently reported an excess in fifteen years
of \fermi\ Large Area Telescope (LAT) data over $|\ell|\le60^\circ$
and $10^\circ\le|b|\le60^\circ$. Its spectrum peaks near 20\,GeV and is
consistent with zero below 2\,GeV, while its radial profile is compatible
with annihilation in a Navarro--Frenk--White (NFW) halo. An interpretation
in terms of annihilation into $b\bar b$ implies a particle mass of
approximately 0.5--0.8\,TeV and a cross section of
$(5$--$8)\times10^{-25}\,\mathrm{cm^3\,s^{-1}}$. Models with
velocity-dependent or resonant annihilation have been proposed to reconcile these
values with constraints from other targets
\cite{Jho:2025iah,Murayama:2025ihg}. The high cross-section is in conflict with Galactic Center constraints \cite{Abazajian:2020tww}, which are imposed on the same halo, so that dynamical differences in the halo cannot reconcile the halo signal with the Galactic Center constraints. These Galactic Center constraints are soon to be updated \cite{AbazajianInPrep}.  Cholis, Hu and Zhong \cite{Cholis:2026rba} attribute the excess
to previously unmodeled structure in the \fermi\ bubbles. Their
model combines updated flat bubbles, a tilted pair of ``cocoons,''
and ``inner bubbles.'' They conclude that these components remove
the need for halo-like emission as bright as that reported by
Totani, while allowing a fainter dark matter contribution.

The possibility that this excess and a high-energy nuclear recoil in
LUX-ZEPLIN (LZ) have a common dark matter origin has attracted further
interest. LZ reported one event consistent with a recoil energy of
$248\pm23\,\mathrm{(stat)}\pm23\,\mathrm{(sys)}$\,keV and a global
background-only significance of $2.6\sigma$ \cite{LZ:2026axp}. 
In particular, 
Yamashita \cite{Yamashita:2026ump} proposed a 420\,GeV dark photon that
annihilates into $W^+W^-$ in the Galactic halo and scatters inelastically in xenon, with a mass splitting near 316\,keV. 
Very significantly, the LZ event favors dark matter masses above approximately 200\,GeV \cite{LZ:2026axp}, so that the nuclear-recoil and gamma-ray observations each independently indicate dark matter at the several-hundred-GeV to TeV mass scale. 
Therefore, whether the gamma-ray excess has the spectrum and spatial distribution expected of halo annihilation, which we test here, bears directly on whether this convergence of the two indications holds.

The uncertain morphology of the large Galactic structures is important
for this assessment. 
The \fermi\ bubbles contain brightness variations
within their boundaries \cite{Su:2010qj,Fermi-LAT:2014sfa,Fermi-LAT:2017opo}, and their emission overlaps the halo search region. 
Radio and X-ray observations also motivate templates for Loop~I and the eROSITA bubbles \cite{Wolleben:2007pq,Predehl:2020kyq}. 
Varying the cosmic-ray propagation model while keeping these templates fixed leaves their morphological uncertainties untested. 
Earlier halo searches found substantial changes in the inferred emission with the diffuse model \cite{Chang:2018bpt}.
An independent examination of Totani's excess by Foster \cite{Foster:2026cern} also reports foreground dependence and differences between the two hemispheres. 
These results make it important to fit the candidate signal together with alternative descriptions of the extended foreground emission.

A further uncertainty arises when the \fermi\ bubbles template is constructed from the gamma-ray residuals. 
Totani derives this template at 4.3\,GeV after subtracting the other fitted components \cite{Totani:2025fxx}. 
Unmodeled emission with a halo-like distribution can remain in the residual map and subsequently be fitted as part of the foreground at other energies. 
An injection test with that map held fixed cannot measure the resulting bias. 
We therefore rebuild the map after adding a signal and also analyze simulated skies without an injected halo. 
Together, these tests assess how the template construction affects signal recovery and whether foreground mismatch can produce an apparent excess.

We examine these effects with 128 GALPROP models in two component
decompositions, the A2020 configuration, and six additional
GALPROP v57 models. We vary the \fermi\ bubbles, Loop~I, and
eROSITA bubbles templates, and use simulated skies to distinguish
signal recovery from foreground-induced biases. None of the
tested \fermi\ bubbles templates is assumed to describe the
true emission.

After reproducing Totani's original $10^\circ$ analysis, we
repeat foreground selection and signal inference over
5.6--130\,GeV with finer spatial bins and the point-spread
function included. We also fit an NFW component alongside the
released substructured \fermi\ bubbles templates of
Ref.~\cite{Cholis:2026rba}. These comparisons test how the
inferred spectrum and spatial distribution depend on the
foreground description, and what that dependence implies for
the proposed dark matter connection to LZ.

Section~\ref{sec:data} describes the observations, instrument response, and emission models, followed by the statistical analysis in Sec.~\ref{sec:method}. Section~\ref{sec:results} presents the finer-binning fits, annihilation-spectrum comparisons, and emission outside the \fermi\ bubbles footprint. 
We examine the spatial consistency of the excess and the foreground tests in 
Sec.~\ref{sec:morphology}, then discuss the interpretation and summarize
our conclusions in Sec.~\ref{sec:discussion}. 
The appendices give the data-reduction and response checks (Appendix~\ref{app:response}), diffuse-model construction (Appendix~\ref{app:diffuse}), and reproduction of the original coarse-binning analysis (Appendix~\ref{app:coarse}). Simulation and numerical checks are described in Appendix~\ref{app:simulations}, with additional eROSITA reconstruction and simulated-halo comparisons in Appendix~\ref{app:morphology}. 
Appendix~\ref{app:fullrange} contains the full-energy-range fits and spectral controls.

\section{Observations and emission models}
\label{sec:data}
\label{sec:templates}

\subsection{Data selection and angular response}
\label{sec:observations}

We use fifteen years of Pass~8 ULTRACLEAN events (mission weeks
009--791), combining front and back conversions \cite{Atwood:2009ez}.
The fitted region spans $|\ell|\le60^\circ$, $10^\circ\le|b|\le60^\circ$.
Nine logarithmic bins cover 1.16--130\,GeV, with centers at 1.5, 2.6,
4.3, 7.3, 12.3, 20.8, 35, 59, and 100\,GeV. Counts and exposure are
first mapped on a $0.125^\circ$ Galactic plate-carr\'ee grid. The
likelihood uses counts summed into cells of width $10^\circ$,
$2^\circ$, or $1^\circ$ (120, 3000, or 12,000 cells). Identical
extended-source masks are applied to data and templates.
Appendix~\ref{app:response} gives the reduction settings.

For the $2^\circ$ and $1^\circ$ fits, we convolve expected-count templates with the \code{P8R3\_ULTRACLEAN\_V3} point-spread function (PSF) before spatial binning. 
Source kernels are centered on catalog positions, not pixel centers. 
Totani's residual \fermi\ bubbles templates already contain instrumental blurring and are not convolved again. 
They are constructed with the original $10^\circ$ fit and projected onto the finer cells. 
Observed counts remain unsmoothed.

Our final comparisons use the six-bin range denoted 5.6--130\,GeV.
In these bins, the energy-averaged 68\% PSF containment radius is no
larger than $0.25^\circ$. 
We repeat foreground selection and signal inference over this range and retain the full-range fits as a separate sensitivity check. 
Comparisons with \code{gtsrcmaps} and calculations on an extended spatial grid support the response treatment in the restricted range (Appendix~\ref{app:response}). 
To assess sensitivity to spatial binning, we compare fitted amplitudes and the likelihood improvements between models fitted at the same binning. 
Absolute likelihoods from different data partitions are not compared.

\subsection{Diffuse emission and cataloged sources}
\label{sec:diffuse}

The diffuse library contains the 128 GALPROP models released by the LAT Collaboration \cite{Ackermann:2012pya}, varying cosmic-ray sources,
propagation dimensions, gas spin temperature, and dust correction.
Each model is fitted in two decompositions. The single-gas
decomposition has one normalization per energy bin for
gas-correlated emission and another for inverse Compton
emission. The five-ring decomposition has five independently
normalized gas groups and three inverse Compton components
associated with optical, infrared, and microwave target
photons. The two decompositions therefore differ in both their
gas and inverse Compton freedom. Together with A2020
\cite{Abazajian:2020tww}, they give 257 configurations, where
a configuration denotes a diffuse model and its decomposition.

Six additional GALPROP v57 models (Galp21) combine three source distributions with the F98 and R12 radiation fields \cite{Porter:2021tlr,Porter:2017vaa}. 
Two inverse Compton treatments and two decompositions give 24 configurations (Appendix~\ref{app:galp21}). 
The LAT model \code{gll\_iem\_v07} is used only as a reference because its residual-derived emission overlaps our \fermi\ bubbles and Loop~I templates \cite{Fermi-LAT:2019yla,Ballet:2023qzs}. 

We construct the catalog-source contribution from 4FGL-DR4,
which covers fourteen of the fifteen years in our data
selection \cite{Ballet:2023qzs}. The free normalization in each
energy bin adjusts the combined source flux, but does not
independently account for source-to-source variability. At
$|b|\ge10^\circ$, the catalog sources predict 10\%--18\% of the
observed counts across the bins from 1.16 to 130\,GeV, including
11\% at 20.8\,GeV. Above a few GeV, source-flux errors mainly
affect localized regions, although this does not by itself
establish that their effect on the fitted NFW amplitude is
negligible.

Extended sources are masked within twice their semi-major axes, removing approximately 1\% of the fitted-region pixels. 
Isotropic emission is always included. 
Solar inverse Compton emission is varied with the other extended components.

\subsection{\fermi\ bubbles, Loop~I, and eROSITA bubbles templates}
\label{sec:structures}

We compare the flat \fermi\ bubbles template, with uniform surface brightness inside the 2014 mask \cite{Fermi-LAT:2014sfa}, the structured \fermi\ bubbles template from the 2017 map \cite{Fermi-LAT:2017opo}, and Totani's residual-derived templates \cite{Totani:2025fxx}. 
The structured template comes from a spectral-component decomposition of 1--10\,GeV data and may contain emission from other components. 
We abbreviate these descriptions as ``flat,'' ``structured,'' and ``residual'' in tables and comparisons.
We also test the substructured model of Ref.~\cite{Cholis:2026rba}, using its public templates for the flat bubbles with enlarged boundaries, the cocoons, and the inner bubbles, each with an independent normalization per energy bin (Sec.~\ref{sec:chz}).

Totani's residual templates are constructed by fitting the 4.3\,GeV
data over the full $120^\circ\times120^\circ$ square, including the
disk. 
The fit contains the diffuse model, flat \fermi\ bubbles
 template, Loop~I shells, isotropic emission, point sources, and the
original NFW-$\rho^{2.5}$ inner-Galaxy template. Subtracting all fitted
components except the \fermi\ bubbles leaves their fitted emission
plus the residual. After exposure correction, masking, and $1^\circ$
Gaussian smoothing, the positive and negative parts become separate
templates with independent normalizations per energy bin. They are
rebuilt for every diffuse configuration, allowing signal present at
the construction energy to enter the inferred \fermi\ bubbles
morphology.

For Loop~I, we use the Wolleben shells \cite{Wolleben:2007pq}, the destriped Haslam 408\,MHz map restricted to the North Polar Spur \cite{Remazeilles:2014mba}, the corresponding raw Haslam map, or the shells combined with the restricted Haslam component. 
For the eROSITA bubbles, we adopt the geometric boundaries
of Ref.~\cite{Zhang:2024mcd} and compare uniform surface
brightness with weights reconstructed from a 0.6--1.0\,keV
X-ray map based on Ref.~\cite{Predehl:2020kyq}.
The northern and southern lobes have independent normalizations, either over their full footprint or only outside the flat \fermi\ bubbles mask. 
The brightness-weighted eROSITA bubbles template uses native survey data in the West. 
Its eastern half is reconstructed from a published color image and calibrated against the western data. 
This reconstruction supplies 43\%--47\% of the template weight in our fitted region.
Appendix~\ref{app:erosita} describes the associated checks and uncertainties.

\subsection{NFW template}
\label{sec:halo}

We adopt the halo profile used by Totani,
\begin{align}
 \rho(r)&=\rho_s(r/r_s)^{-1}(1+r/r_s)^{-2},\\
 J(\ell,b)&=\int_0^{200\,\mathrm{kpc}}\rho^2[r(s,\ell,b)]\,ds,
 \label{eq:jmap}
\end{align}
with $r_s=21$\,kpc, $R_\odot=8$\,kpc, and
$\rho_\odot=0.42\,\mathrm{GeV\,cm^{-3}}$ \cite{Totani:2025fxx}.
We fit an independent amplitude in every energy bin and call the
result the NFW component, without assigning it a dark matter origin.
The fitted coefficient multiplies $J$ to give the differential
intensity, approximated as constant within each energy bin.
The coefficient has units of
$\mathrm{cm^3\,s^{-1}\,GeV^{-2}\,MeV^{-1}\,sr^{-1}}$.
Unless otherwise stated, quoted counts refer to the NFW
contribution in the bin centered at 20.8\,GeV over the full
fitted region. Changing the adopted density normalization
rescales the fitted coefficient and its conversion to an
annihilation cross section, but leaves the predicted counts
and likelihood unchanged.

\begin{figure*}[t]
\centering
\includegraphics[width=\textwidth]{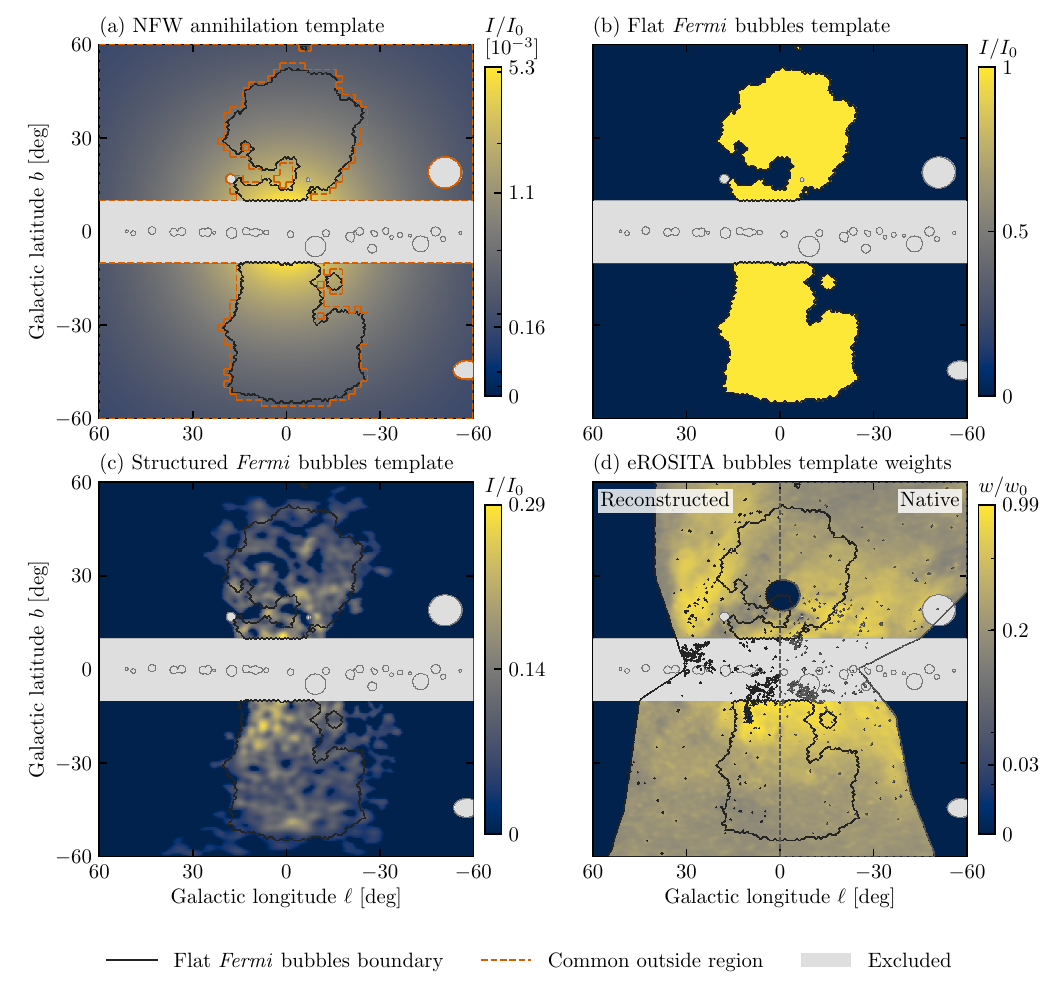}
\caption{Shown here are the spatial templates whose fitted amplitudes are
compared in this work, on the $120^\circ\times120^\circ$ analysis grid
before exposure weighting and PSF convolution: (a) the NFW annihilation
template, $J(\ell,b)$ of Eq.~(\ref{eq:jmap}); (b) the flat \fermi\ bubbles
template \cite{Fermi-LAT:2014sfa}; (c) the structured \fermi\ bubbles
template \cite{Fermi-LAT:2017opo}; and (d) the
X-ray-brightness-weighted eROSITA bubbles template
\cite{Predehl:2020kyq,Zhang:2024mcd}. The NFW template overlaps the
\fermi\ bubbles over much of the fitted region at $|b|\lesssim30^\circ$,
and the eastern ($\ell>0$) half of the eROSITA bubbles template, which
reduces the gamma-ray asymmetry in Sec.~\ref{sec:erositafreedom}, is a
reconstruction rather than native survey data. Each map is divided by
its maximum over the full template, including the masked plane, so that
the NFW template within the fitted region reaches only $5.3\times10^{-3}$
of its central value; the color scales of (a) and (d) are stretched for
display, and the northern and southern halves of the eROSITA template
retain their independent normalizations. The solid (black) line is the
boundary of the flat \fermi\ bubbles, the dashed (orange) line in (a)
encloses the region outside the bubbles used in Sec.~\ref{sec:outside},
and the dashed (gray) line in (d) separates the native western from the
reconstructed eastern eROSITA data. Light gray marks the excluded plane,
$|b|<10^\circ$, and the 4FGL-DR4 extended sources masked within twice
their semi-major axes \cite{Ballet:2023qzs}. The dark disk at
$(\ell,b)\approx(-1^\circ,+24^\circ)$ in (d) is the region around
Sco~X-1 outside the eROSITA coverage, where the template weight is zero.}
\label{fig:template_overview}
\end{figure*}

Figure~\ref{fig:template_overview} shows the NFW and bubble templates together with the masks that define the fitted region.
\clearpage
\section{Statistical analysis}
\label{sec:method}

\subsection{Template likelihood and significance}
\label{sec:likelihood}

For energy bin $e$ and spatial cell $p$, the mean count is
\begin{equation}
 \mu_{ep}=a_e H_{ep}+\sum_j\eta_{ej}T_{ejp},
 \label{eq:counts}
\end{equation}
where $H$ is the NFW template with amplitude $a_e$ and $T_j$ are the foreground templates with normalizations $\eta_{ej}$, all expressed in expected-count units. 
Here, $L_e$ is the Poisson likelihood of the counts in bin $e$, and $\boldsymbol\eta_e=(\eta_{e1},\eta_{e2},\ldots)$ denotes the foreground normalizations in that bin. 
At each trial value of $a_e$, we maximize $L_e$ over $\boldsymbol\eta_e$.
The spectral fits allow either sign of $a_e$, provided that the total
expected count remains positive in every cell. We quote 68\% profile
intervals and describe the numerical refinement and fixed-template
injection checks in Appendix~\ref{app:estimator}.

To test for positive emission, we constrain the NFW amplitude to be
nonnegative and compare the fit with a model that omits this component
\cite{Macias:2016nev},
\begin{align}
 q_e^+&=2\left[\max_{a_e\ge0,\boldsymbol\eta_e}\ln L_e
       -\max_{\boldsymbol\eta_e}\ln L_e(0,\boldsymbol\eta_e)\right],\\
 Q^+&=\sum_{e=1}^{K}q_e^+ .
 \label{eq:qplus}
\end{align}
A bin whose signed best-fit amplitude is negative contributes
zero to $Q^+$. For fixed templates, independent energy bins,
and regular nuisance parameters, the asymptotic null
distribution is the binomial mixture
\begin{equation}
 f_K(q)=2^{-K}\delta(q)+2^{-K}\sum_{j=1}^{K}\binom{K}{j}f_{\chi_j^2}(q).
 \label{eq:mixture}
\end{equation}
We convert the upper-tail probability $p_K$ to a Gaussian-equivalent
significance using the convention of Ref.~\cite{Macias:2016nev},
$Z_{2018}=\Phi^{-1}(1-p_K/2)$, where $\Phi$ is the standard-normal
cumulative distribution \cite{Cowan:2010js}. The restricted and full
energy ranges have $K=6$ and $K=9$ tested amplitudes, respectively.
The coarse-binning appendices also report the signed single-bin
quantity $\mathrm{sign}(a_e)\sqrt{\mathrm{TS}_e}$, which is distinct
from this combined significance.

Each significance applies to the foreground configuration in which it
was calculated. The configurations are alternative descriptions of the
same data, so we do not multiply their likelihoods or interpret their
relative frequencies as probabilities. Medians and ranges across the
model library describe the foreground dependence of the fits, not
confidence intervals. Our fixed-configuration simulations test the
corresponding conditional statistic. They do not calibrate a global
dark matter significance that includes every model choice.

\subsection{Foreground selection and model comparisons}
\label{sec:selection}
\label{sec:ranking}

For each complete foreground configuration, we compare fits with and
without the NFW component while keeping the same nuisance freedoms.
We then compare the \fermi\ bubbles templates with the NFW component
included. This second comparison is necessary because the preferred
foreground template can change when additional emission is allowed.

We also select extended structures by adding them one at a time to a
background-only fit. For a candidate with $\Delta p$ additional
binwise normalizations, we require an improvement in $-2\ln\mathcal L$
larger than its Bayesian information criterion (BIC) cost,
$\Delta p\ln n$, with $n=K N_{\rm cell}$ \cite{Schwarz:1978bic}.
A candidate is eligible if it clears this threshold in at least 80\%
of the diffuse configurations. Of the eligible candidates, we add the
one that gives the largest improvement in the greatest number of
configurations, remove the alternative templates for that structure,
and repeat. We run this forward selection separately for the two
diffuse decompositions, with and without Totani's residual templates
among the candidates. We also record the sequence selected by each
configuration individually.

The final shared selection uses 16 balanced 2012 models in each
decomposition, together with A2020. Galp21 models are fitted with the
selected templates and undergo their own individual selections, but
do not contribute to the shared choice. Table~\ref{tab:res} compares
six representative diffuse configurations with identical accompanying
foreground templates. We do not assign Gaussian significances to BIC
scores or to likelihood differences between nonnested \fermi\ bubbles
models.

\subsection{Spectral and morphological tests}
\label{sec:gates}

We test two benchmark annihilation spectra from PPPC4DMID \cite{Cirelli:2010xx}: a 500\,GeV particle annihilating into $b\bar b$, representative of the interpretation of Totani's excess, and a 420\,GeV particle annihilating into $W^+W^-$, as proposed for the connection to the LZ event \cite{Yamashita:2026ump}. 
For each spectrum, we fit a single signed coefficient that scales the fixed spectral shape across all $K$ energy bins, refitting the foreground normalizations, and compare it with $K$ independent signed NFW amplitudes. 
If the shape is correctly specified and the parameters are identifiable and interior, the asymptotic reference for this likelihood difference is $\chi^2_{K-1}$.
Therefore, the test assesses each benchmark spectrum under a given foreground model, rather than determining the particle mass or annihilation channel, and we do not fit a joint gamma-ray and direct-detection likelihood.

To test the spatial distribution, we compare one NFW normalization across the fitted region with four independent spatial quadrant normalizations in each energy bin. 
A common coefficient scales the prescribed NFW intensity pattern, rather than imposing uniform surface brightness.
We divide the intensity template before PSF convolution so that equal coefficients reproduce the full folded template, including spillover across quadrant boundaries. We test the standard quadrants, divided north/south and east/west, and a partition rotated by $45^\circ$. 
Each fixed partition adds three parameters per bin. 
We denote by $Q_6$ the sum of $\dl$ between the common and regional fits over the six bins from 5.6 to 130\,GeV.
Simulations drawn from the fitted common-normalization model calibrate both tests and the statistic obtained by taking the larger of the two.  
Unequal fitted normalizations indicate that the NFW component together with the adopted foreground does not describe the spatial distribution adequately. 
A spherical contribution may still be present alongside asymmetric residual emission.

\subsection{Simulated skies}
\label{sec:simmethod}

We draw Poisson realizations from representative foreground fits, with and without an added NFW signal. 
One injected spectrum follows our reproduction of Totani's excess and contributes 6412 counts at 20.8\,GeV. We also inject an exactly specified $b\bar b$ spectrum.
In the residual-template construction tests, we rebuild
Totani's \fermi\ bubbles templates from each realization's
4.3\,GeV pixels and compare the results with controls in
which the templates are held fixed. The original construction
tests retain the observed counts in the 24 disk cells and
are therefore conditional on the disk data.

We repeat foreground selection in each simulated sky for
the selection experiments, but keep the configuration fixed
in the final 50-sky calibration ensembles. The quadrant
tests use 200-sky reference ensembles at $2^\circ$ and
50-sky ensembles at $1^\circ$, with any maximization over
partitions repeated in each realization. The generating
configurations are specified in
Secs.~\ref{sec:quadrantfinal} and \ref{sec:chz}. These
ensembles test recovery, bias, and the central null
distributions; they do not resolve the very small tail
probabilities associated with the largest analytic
significances.

Appendix~\ref{app:simulations} describes the separate experiments. 

\section{NFW-correlated emission}
\label{sec:results}
\label{sec:resp}

\subsection{Spatial binning and foreground dependence}
\label{sec:twodeg}

Totani's original configuration, his baseline diffuse model with the residual-derived \fermi\ bubbles templates, gives $6412\,[5904,6898]$ counts at 20.8\,GeV and $\mathrm{TS}=172$.
The amplitudes are positive in all four quadrants.
Replacing only the residual \fermi\ bubbles templates gives 5006 counts with the flat template and 1575 with the structured template.
The original $10^\circ$ fit is therefore sensitive to this choice.
Appendix~\ref{sec:repro} gives the detailed reproduction, and
Fig.~\ref{fig:ranking} shows the dependence across the diffuse
configurations at all three binnings.

The finer-binning fits, with the response included, retain a positive
NFW-correlated component for all three \fermi\ bubbles descriptions
(Table~\ref{tab:res}). 
With Totani's single-gas diffuse model and the structured \fermi\ bubbles template, the amplitude rises to $3158\,[2598,3718]$ counts at $2^\circ$ and
$3527\,[2982,4073]$ at $1^\circ$. 
The flat and residual templates give larger amplitudes at both binnings. 
The much smaller value obtained with the structured template in the original $10^\circ$ fit thus does not establish an absence of NFW-correlated emission (Fig.~\ref{fig:ranking}, lower rows).

\begin{table*}[t]
\caption{Tabulated here are the response-folded NFW fits for six
representative diffuse configurations, each fitted with the structured,
flat, and residual-derived (``construction'') \fermi\ bubbles templates
and identical accompanying foregrounds (uniform north/south eROSITA
bubbles templates, Haslam Loop~I, solar inverse Compton emission,
isotropic emission, and catalog sources). An NFW-correlated component
is present in every configuration at both binnings, with
$Z_{2018}=4.9$--32.8. However, its amplitude depends on the foreground
model: in every configuration, the amplitude increases from the
structured to the flat to the residual-derived templates, which give a
factor of 1.4--2.1 more counts than the structured template. Here,
$A_{20.8}$ is the number of NFW counts at 20.8\,GeV, with 68\% profile
intervals in brackets, for $2^\circ$ and $1^\circ$ cells. 
$Z_{2018}$ is the Gaussian-equivalent significance of
$Q^+$ over 5.6--130\,GeV, using the convention of
Ref.~\cite{Macias:2016nev}. It is conditional on the
foreground configuration, not a global dark matter
significance. The labels 1-gas and 5-ring denote the
decompositions defined in Sec.~\ref{sec:diffuse}. 
``Totani'' marks his baseline diffuse model, and ``best'' the best-fitting configuration of each decomposition selected without residual-derived templates, which also generate the simulated skies.}
\label{tab:res}
\centering
\fontsize{8.7}{10.2}\selectfont
\setlength{\tabcolsep}{3pt}
\begin{ruledtabular}
\begin{tabular}{llrrrr}
diffuse configuration & bubbles & $A_{20.8}$, $2^\circ$ & $A_{20.8}$, $1^\circ$ & $Z$, $2^\circ$ & $Z$, $1^\circ$\\
\hline
\code{SLZ6R30T150C2} 1-gas (Totani) & structured & 3158 [2598, 3718] & 3527 [2982, 4073] & 7.3 & 9.3\\
 & flat & 4932 [4431, 5434] & 5021 [4528, 5517] & 17.7 & 18.7\\
 & construction & 6759 [6295, 7224] & 6896 [6439, 7355] & 28.8 & 30.3\\
 \code{SOZ6R30T150C5} 1-gas (best 1-gas)& structured & 4272 [3750, 4794] & 4565 [4058, 5074] & 13.8 & 15.9\\
 & flat & 6153 [5685, 6622] & 6200 [5741, 6661] & 28.5 & 29.7\\
 & construction & 6445 [5996, 6896] & 6617 [6177, 7059] & 30.7 & 32.8\\
 \code{SLZ6R30T150C2} 5-ring  & structured & 4576 [4019, 5098] & 4869 [4344, 5376] & 7.9 & 9.4\\
 & flat & 6294 [5442, 6872] & 6345 [5538, 6909] & 9.9 & 10.9\\
 & construction & 7357 [6800, 7991] & 7512 [6963, 8132] & 13.4 & 14.8\\
\code{SSZ10R30T150C5} 5-ring (best 5-ring) & structured & 4441 [3391, 5144] & 4815 [3928, 5511] & 4.9 & 6.1\\
 & flat & 5679 [4484, 6704] & 5916 [4738, 6825] & 6.1 & 7.0\\
 & construction & 7333 [6576, 8051] & 7444 [6704, 8101] & 11.6 & 12.6\\
Galp21 SA0\_R12 1-gas & structured & 3953 [3407, 4500] & 4336 [3806, 4868] & 10.4 & 12.8\\
 & flat & 5679 [5201, 6156] & 5800 [5321, 6280] & 23.0 & 24.3\\
 & construction & 6544 [6089, 7000] & 6768 [6323, 7214] & 29.8 & 32.1\\
A2020 & structured & 5761 [5194, 6329] & 6125 [5576, 6676] & 14.1 & 15.8\\
 & flat & 7404 [6914, 7888] & 7521 [7015, 8030] & 19.6 & 20.3\\
 & construction & 8538 [8076, 9005] & 8689 [8220, 9159] & 20.3 & 22.2\\
\end{tabular}
\end{ruledtabular}
\end{table*}

Across the six representative configurations, refining the spatial
bins from $2^\circ$ to $1^\circ$ changes the NFW amplitude by 1--4\%
with the flat or residual \fermi\ bubbles templates and by 6--12\%
with the structured template. The profile intervals overlap. With the
NFW component included, the flat template fits better than the
structured template by 140--380 in $-2\ln\mathcal L$ over
5.6--130\,GeV, and the differences at the two binnings agree within
50 units. The residual templates improve the fit over the flat template
by a further 670--1020, although they were constructed from the data
being fitted. In fits without the NFW component, the preferred choice
between the flat and structured templates changes with the diffuse
configuration.

Table~\ref{tab:sel} presents the results after repeating the
shared background-only model selection over
5.6--130\,GeV and then fitting the NFW component. The
single-gas decomposition selects the structured \fermi\
bubbles template, whereas the five-ring decomposition
selects the flat \fermi\ bubbles template and uniform
eROSITA bubbles templates outside the 2014 \fermi\ bubbles
mask. This selection is distinct from the fixed-companion
comparison in Table~\ref{tab:res}, where the \fermi\
bubbles alternatives are compared with NFW included.

The median
NFW amplitudes over the 17 configurations in each decomposition are
4097 and 4454 counts for the single-gas fits at $2^\circ$ and
$1^\circ$, respectively, and 6154 and 6281 counts for the five-ring
fits. Individual configurations span a wider range, as illustrated in
Table~\ref{tab:res}. 
The six representative configurations include Totani's baseline, \code{SLZ6R30T150C2}, in both decompositions; the best-fitting single-gas and five-ring configurations selected without residual-derived templates, \code{SOZ6R30T150C5} 1-gas and \code{SSZ10R30T150C5} 5-ring, which also provide the generating diffuse models for the corresponding simulation tests; A2020; and Galp21 SA0\_R12.
The two best-fitting configurations also
provide the generating diffuse models for the corresponding
simulation tests.

Note that $Z_{2018}$ follows the two-sided convention of Ref.~\cite{Macias:2016nev}, so that the one-sided discovery significance conventional in particle physics \cite{Cowan:2010js} is lower by at most 0.15$\sigma$ over the range in Table~\ref{tab:res}; 
in either convention, these are local significances for a fixed foreground model, evaluated from the asymptotic distribution in Eq.~(\ref{eq:mixture}), and include no trials factor for the many foreground models tested.

\begin{table}[tbp]
\caption{Tabulated here are the reselected, response-folded fits
without the residual-derived \fermi\ bubbles templates. Over all nine
energy bins, both decompositions select the flat \fermi\ bubbles
template; over 5.6--130\,GeV, the single-gas decomposition selects the
structured template instead (bold), lowering the NFW counts and $\sv$.
Counts (at 20.8\,GeV), $Z_{2018}$, and $\sv$ (for 500\,GeV $b\bar b$,
in $10^{-25}\,\mathrm{cm^3\,s^{-1}}$) are medians over the 17
configurations in each decomposition. eU and eO are uniform
north/south eROSITA bubbles templates over the full or outer
footprint, H is Haslam emission, and IC is solar inverse Compton
emission.}
\label{tab:sel}
\centering
\fontsize{8}{9.5}\selectfont
\setlength{\tabcolsep}{3pt}
\begin{ruledtabular}
\begin{tabular}{lllrrr}
cells, bins & gas & selected & counts & $Z$ & $\sv$\\
\hline
$2^\circ$, all & 1-gas & flat, eU, H, IC & 5735 & 28.7 & $+5.5$\\
 & 5-ring & flat, eU, H, IC & 6205 & 11.5 & $+6.0$\\
$1^\circ$, all & 1-gas & flat, eU, H, IC & 5821 & 30.0 & $+5.7$\\
 & 5-ring & flat, eU, H, IC & 6335 & 12.9 & $+6.4$\\
$2^\circ$, 5.6--130 & 1-gas & \textbf{structured}, eU, H & 4097 & 12.6 & $+3.1$\\
 & 5-ring & flat, eO & 6154 & 11.2 & $+6.1$\\
$1^\circ$, 5.6--130 & 1-gas & \textbf{structured}, eU, H & 4454 & 14.8 & $+3.5$\\
 & 5-ring & flat, eO & 6281 & 12.0 & $+6.2$\\
\end{tabular}
\end{ruledtabular}
\end{table}

The Galp21 fits also retain the component. All 12 single-gas
configurations select the structured \fermi\ bubbles template in
their individual 5.6--130\,GeV selections and give 3600--4400 NFW
counts at $2^\circ$. Including all nine energy bins instead selects
the flat \fermi\ bubbles template in both diffuse decompositions
and all 24 Galp21 configurations. Thus, the three lowest energy bins
can determine the foreground choice even when the quoted signal
amplitude is measured near 20\,GeV. These full-range fits are given in
Appendix~\ref{app:fullrange}.

\subsection{Annihilation spectra and the LZ benchmark}
\label{sec:dm}

The proposed LZ interpretation uses the gamma-ray spectrum from a
420\,GeV particle annihilating into $W^+W^-$
\cite{Yamashita:2026ump}. We fit its normalization through the binwise
likelihoods and compare it with the independently fitted NFW amplitudes.
This tests the spectral shape under each foreground model, without
imposing the rate predicted by the complete particle scenario or using
the LZ likelihood.

\begin{figure*}[t]
\centering
\includegraphics[width=0.94\textwidth]{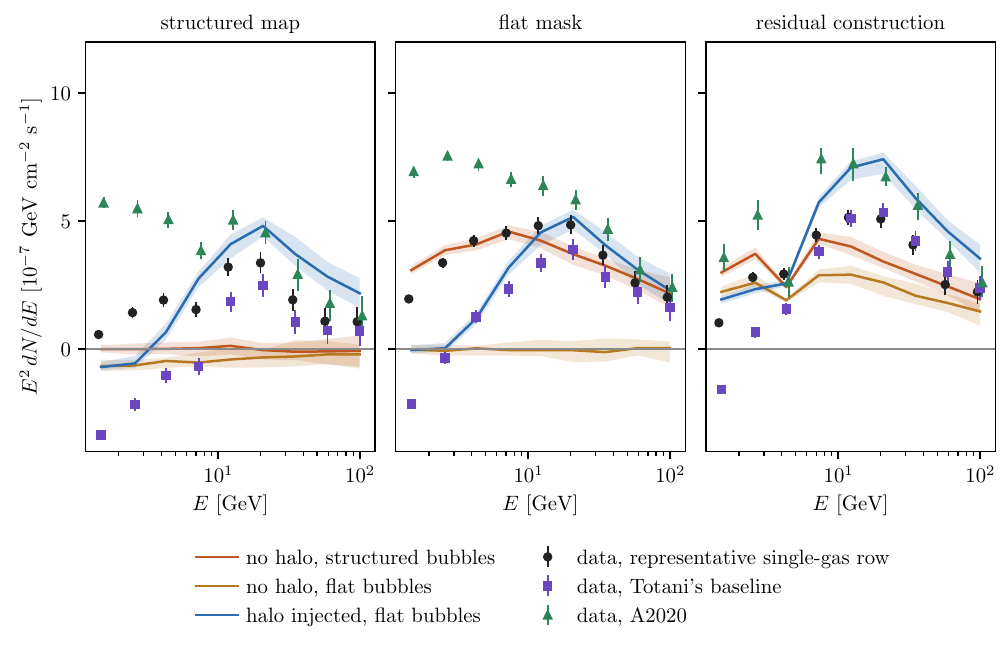}
\caption{Shown here are the response-folded NFW spectra in $2^\circ$
cells fitted with the structured, flat, and residual-derived \fermi\
bubbles templates (left to right).
When the generating and fitted \fermi\ bubbles
descriptions differ, no-halo simulations can yield
a spurious NFW component with a broad, bubbles-like
spectrum. These examples do not reproduce the
12--21\,GeV feature retained in the data with the
structured template. Rebuilding the residual-derived
templates for each realization also produces
substantial apparent NFW emission in no-halo skies
and suppresses the recovered contribution at the
4.3\,GeV construction energy. The (black) circles, (purple) squares, and
(green) triangles are the data for the representative single-gas
configuration, Totani's baseline, and A2020, with 68\% profile
intervals. The solid lines and bands are the medians and 16--84\%
ranges of fifty simulated skies without a halo on structured (red) or
flat (orange) \fermi\ bubbles, or with the empirical Totani spectrum
injected on flat \fermi\ bubbles (blue).}
\label{fig:specsims}
\end{figure*}

Figure~\ref{fig:specsims} shows how the fitted spectra change with the
\fermi\ bubbles template. With the structured template, a feature
remains at 12--21\,GeV that is not reproduced by the matched simulations
without an NFW signal. In the representative single-gas configuration,
its 20.8\,GeV amplitude is 4272 counts, above all 50 matched null
simulations, but the spectrum falls more rapidly by 35\,GeV than the
injected Totani spectrum. With the flat \fermi\ bubbles template,
Totani's baseline follows the injected shape more closely. The
representative single-gas fit instead resembles the component produced
by a \fermi\ bubbles mismatch below 10\,GeV and exceeds it near the
peak.

The median fitted 500\,GeV $b\bar b$ coefficients after selection are
$(3.1$--$3.5)\times10^{-25}\,\mathrm{cm^3\,s^{-1}}$ for the single-gas
fits and $(6.1$--$6.2)\times10^{-25}$ for the five-ring fits. Both the
500\,GeV $b\bar b$ and 420\,GeV $W^+W^-$ shapes give acceptable fits
with the flat or residual \fermi\ bubbles templates for Totani's
diffuse model in either decomposition, the second five-ring configuration,
and the Galp21 configuration. Comparing these shapes with independent
bin amplitudes over 5.6--130\,GeV gives conditional $p$-values of
0.2--0.9 for $b\bar b$ and 0.05--0.7 for $W^+W^-$.

Other foreground choices give poorer agreement. With the flat or
residual \fermi\ bubbles templates, A2020 disfavors both shapes
($p\le0.003$); with the structured template, $b\bar b$ gives $p=0.02$. In the representative single-gas configuration,
$W^+W^-$ gives $p\simeq10^{-3}$, while $b\bar b$ gives
$p=0.02$--0.05. With the structured \fermi\ bubbles template, the
single-gas fits give $p$-values from $10^{-6}$ to 0.04 for the two
shapes. The tested $\tau^+\tau^-$ spectrum is disfavored in every
reported configuration. These comparisons apply to the specified
benchmarks, rather than all particle masses or mixtures of annihilation
channels. Appendix~\ref{app:spectralcontrols} gives the full-range
signed coefficients, whose signs can be driven by the lowest energy
bins.

\subsection{Emission outside the \fermi\ bubbles footprint}
\label{sec:outside}

We repeat the fits outside the flat \fermi\ bubbles mask to reduce
the influence of the uncertain brightness distribution inside the
bubbles. This region comprises 2190 $2^\circ$ cells with no mask
coverage and the 8760 $1^\circ$ cells contained within them. Both
binnings therefore use exactly the same sky area, containing 52.5\%
of the NFW template. Table~\ref{tab:outside} reports the counts fitted
within this region, without extrapolating to the full halo.

\begin{table}[tbp]
\caption{Tabulated here are the NFW counts at 20.8\,GeV fitted outside
the flat \fermi\ bubbles mask, a region containing 52.5\% of the NFW
template and identical at both binnings. Each range spans the fits with
the flat, structured, and residual-derived \fermi\ bubbles templates
for a fixed diffuse configuration, and is not a statistical interval.
Outside the bubbles, the choice of bubbles template changes the counts
by at most 24\%, whereas the diffuse configuration changes them by a
factor of $\sim$2.}
\label{tab:outside}
\centering\small
\begin{ruledtabular}
\begin{tabular}{lrr}
diffuse configuration & $2^\circ$ & $1^\circ$\\
\hline
Totani, single-gas & 2430--2800 & 2520--2910\\
Representative single-gas & 3270--3530 & 3440--3700\\
Totani, five-ring & 2710--2780 & 2720--2960\\
Representative five-ring & 2360--2930 & 2510--3070\\
A2020 & 4300--4650 & 4650--5020\\
\end{tabular}
\end{ruledtabular}
\end{table}

Outside the 2014 flat \fermi\ bubbles mask, changing the
\fermi\ bubbles template at fixed diffuse model and spatial
binning shifts the fitted amplitude by at most 570 counts.
The diffuse-model dependence is larger: the fitted
contribution spans 2360--4650 counts at $2^\circ$ and
2510--5020 counts at $1^\circ$, approximately a factor of
two. The nominal six-bin significances are $4$--$15\sigma$
and $4$--$16\sigma$, respectively. Injections in this common
region recover 0.95--1.03 of the input contribution per
energy bin. Extrapolating these fits to the full fitted
region with the NFW morphology would give 70--150\% of
the amplitude in our reproduction of Totani's result.

The inference is more sensitive to the boundary of the
excluded region. The updated boundaries of
Ref.~\cite{Cholis:2026rba} intersect 86 of the 2190
$2^\circ$ cells, containing 13.8\% of the NFW-template
counts in the original outside region. Excluding these
cells lowers the nominal significance with the flat
\fermi\ bubbles template to $1.1$--$7.1\sigma$; for
Totani's baseline, it falls from $7.3\sigma$ to
$1.2\sigma$ (Sec.~\ref{sec:chz}). Thus, the preference
outside the bubbles depends on both the diffuse foreground
and the adopted boundaries.

\begin{figure*}[t]
\centering
\includegraphics[width=\textwidth]{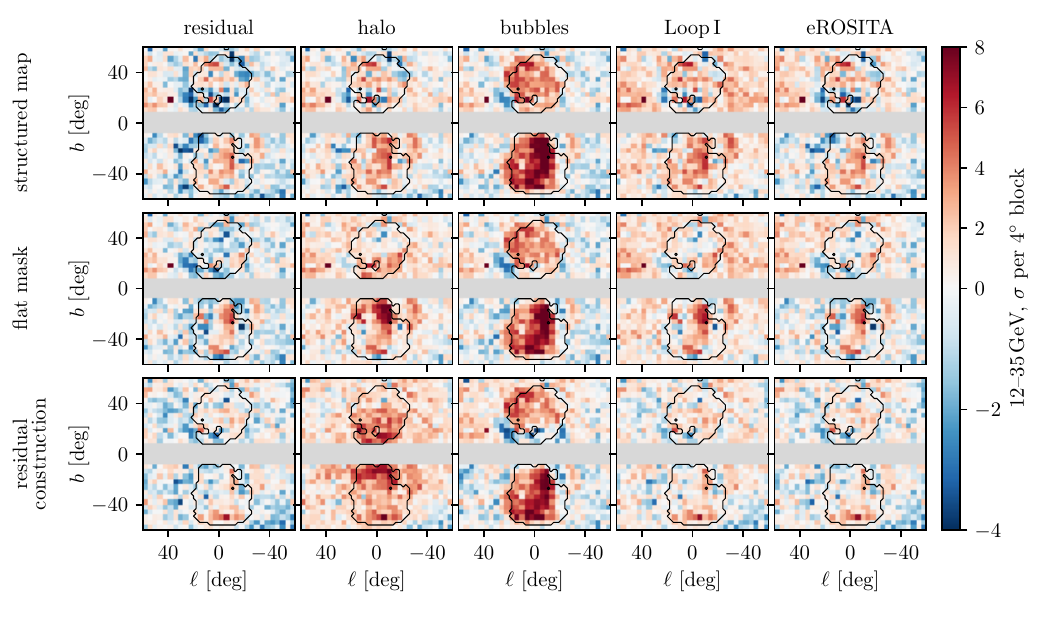}
\caption{Shown here are the residual maps for Totani's baseline
diffuse model, fitted with the PSF in $2^\circ$ cells and displayed
over
$12$--$35\,\mathrm{GeV}$, counts$/\sqrt{\widehat{\mu}}$
per $4^\circ$ block for the structured, flat, and
residual-derived \fermi\ bubbles templates (rows). With the flat and
structured templates, a positive residual remains in the south-western
part of the southern bubble (first column) at $\ell\approx-5^\circ$ to $-10^\circ$,
$b\approx-15^\circ$ to $-30^\circ$, which the residual-derived
templates absorb. The first column is the residual $R=n-\widehat\mu$,
and each later column adds back one fitted component,
$R+\widehat\mu_j$, for the NFW halo, \fermi\ bubbles, Loop~I, and
eROSITA bubbles templates; all maps are divided by
$\sqrt{\widehat\mu}$. Since the residual appears in every
component-added map, these are not independent reconstructions of
the components or calibrated significances. The black outline is the
flat \fermi\ bubbles mask, and gray cells are excluded.}
\label{fig:resid}
\end{figure*}

Inside the flat \fermi\ bubbles mask, the fitted NFW contribution
is more sensitive to the \fermi\ bubbles template. It spans
approximately 600--2000 counts and is consistent with zero in the
five-ring case. With either the flat or the structured template, a
positive residual also remains near $\ell=-5^\circ$ to $-10^\circ$
and $b=-15^\circ$ to $-30^\circ$ (Fig.~\ref{fig:resid}), in the
south-western part of the southern bubble. This feature is distinct
from the south-eastern enhancement of the \fermi\ bubbles reported in
Ref.~\cite{Fermi-LAT:2014sfa} and modeled as the southern cocoon in
Ref.~\cite{Cholis:2026rba}, which is centered near
$(\ell,b)\approx(+8^\circ,-23^\circ)$. Totani's residual templates absorb this
feature more effectively. The component-added residual maps illustrate
where emission remains, but should not be read as independent maps of
the physical components.

\section{Morphology and foreground systematics}
\label{sec:morphology}
\label{sec:erosita}

\subsection{North/south asymmetry and the eROSITA bubbles}
\label{sec:erositafreedom}

Importantly, with the flat \fermi\ bubbles template and uniform eROSITA bubbles
templates, the fitted NFW emission is brighter in the south. Its
south/north count ratio at 20.8\,GeV is approximately 1.4, compared
with 1.05 in the corresponding spherical-signal injections. Allowing
independent northern and southern \fermi\ bubbles normalizations
reduces the ratio to 1.26, but the quadrant fits still favor unequal
NFW amplitudes. The northern and southern spectra have similar shapes
above 12\,GeV, while the south contains an additional soft contribution.
The asymmetry is smaller with A2020, for which the hemispheric spectral
shapes are consistent.

To separate the effects of spatial freedom and X-ray
brightness weighting, we divide both eROSITA bubbles
templates into the same twelve regions: two hemispheres,
two longitude ranges separated at $\ell=0$, and three
latitude bands. Each region has an independent normalization
in every energy bin. The two descriptions therefore have
the same number of free amplitudes, but different intensity
patterns within each region.

\begin{figure*}[t]
\centering
\includegraphics[width=0.9\textwidth]{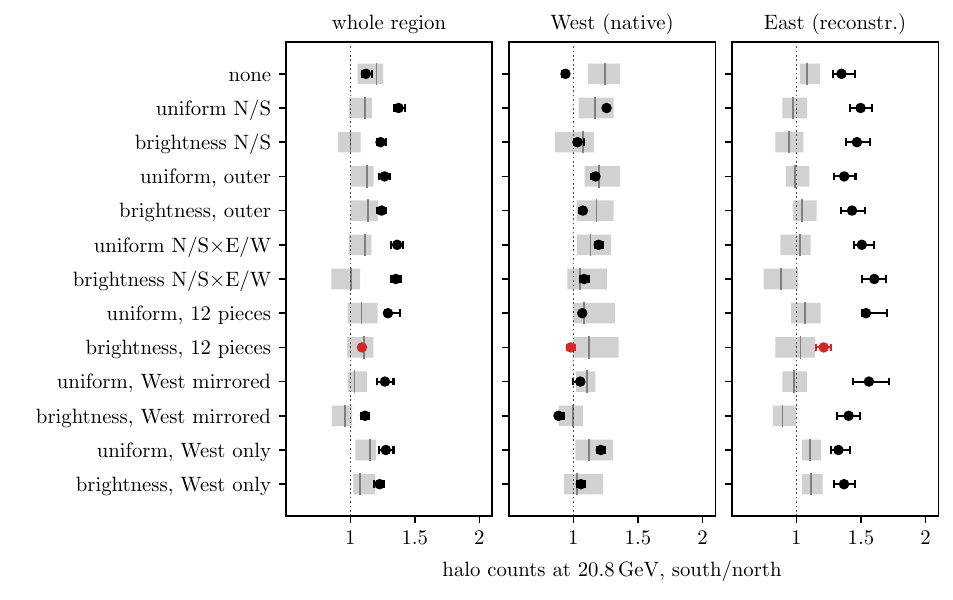}
\caption{
South/north ratios of fitted NFW counts at 20.8\,GeV
for thirteen eROSITA bubbles treatments, including
the case with no eROSITA component. The fits use
the flat \fermi\ bubbles template, the PSF, and
$2^\circ$ cells. Columns show the whole fitted
region, the West with native X-ray data, and the
East with the reconstructed X-ray map. Most
treatments leave a larger departure from the
corresponding spherical-signal injections in
the East than in the West. The twelve
brightness-weighted pieces bring the whole-region ratio close to the
injected value, but still fail the rotated-quadrant test and are not
preferred over twelve uniform pieces when the NFW component is
included. The points and error bars are the median and 16--84\% range
over 34 diffuse configurations, and the gray bands show the
spherical-signal injections.}
\label{fig:erositaflex}
\end{figure*}

Relative to two uniform north/south templates, the twelve
brightness-weighted pieces reduce the median south/north
ratio of fitted NFW counts from 1.37 to 1.09. Twelve
uniform pieces leave a ratio of 1.29
(Fig.~\ref{fig:erositaflex} and
Table~\ref{tab:erositaflex}). The larger reduction
therefore depends on the brightness weights, rather than
simply on the additional normalizations. With those
weights, a common NFW normalization still gives
approximately 4700 counts at 20.8\,GeV with
$\mathrm{TS}=51$, while independent quadrant
normalizations give 4860 counts in total.

The likelihood does not independently favor the X-ray brightness
pattern that restores north/south agreement. Without an NFW component,
twelve brightness-weighted eROSITA bubbles pieces improve
$-2\ln\mathcal L$ by 1278 relative to two uniform lobes, but twelve
uniform pieces already give an improvement of 1196. Most of this gain
therefore comes from the extra spatial freedom. With the NFW component
included, the uniform pieces fit better by a median of 200 over all
nine energy bins, while the two twelve-piece models differ by only
4 over 5.6--130\,GeV. The restricted-range comparison between two-lobe
templates also favors the uniform description, by 150.

The gamma-ray asymmetry is strongest in the East, even when the eastern
eROSITA bubbles template is omitted. The brightness-weighted piece
that reduces it lies in the reconstructed south-eastern region at
$|b|<25^\circ$. This region contains the south-eastern enhancement of
the \fermi\ bubbles \cite{Fermi-LAT:2014sfa}, modeled as the southern
cocoon in Ref.~\cite{Cholis:2026rba}, and the asymmetry can be absorbed
by either an eROSITA bubbles or a \fermi\ bubbles description of this
region (Sec.~\ref{sec:chz}). In separate $10^\circ$ fits, the eastern data favor
the brightness-weighted eROSITA bubbles template, whereas the native
western data favor the uniform template. Alternative reconstructions
preserve the broad brightness contrasts, but do not constrain errors
shared by all of the reconstructions. The result therefore depends
in part on an X-ray morphology that has not been checked against
native eastern survey data.

\subsection{Quadrant likelihood tests and simulated halos}
\label{sec:quadrantfinal}

\begin{figure*}[t]
\centering
\includegraphics[width=\textwidth]{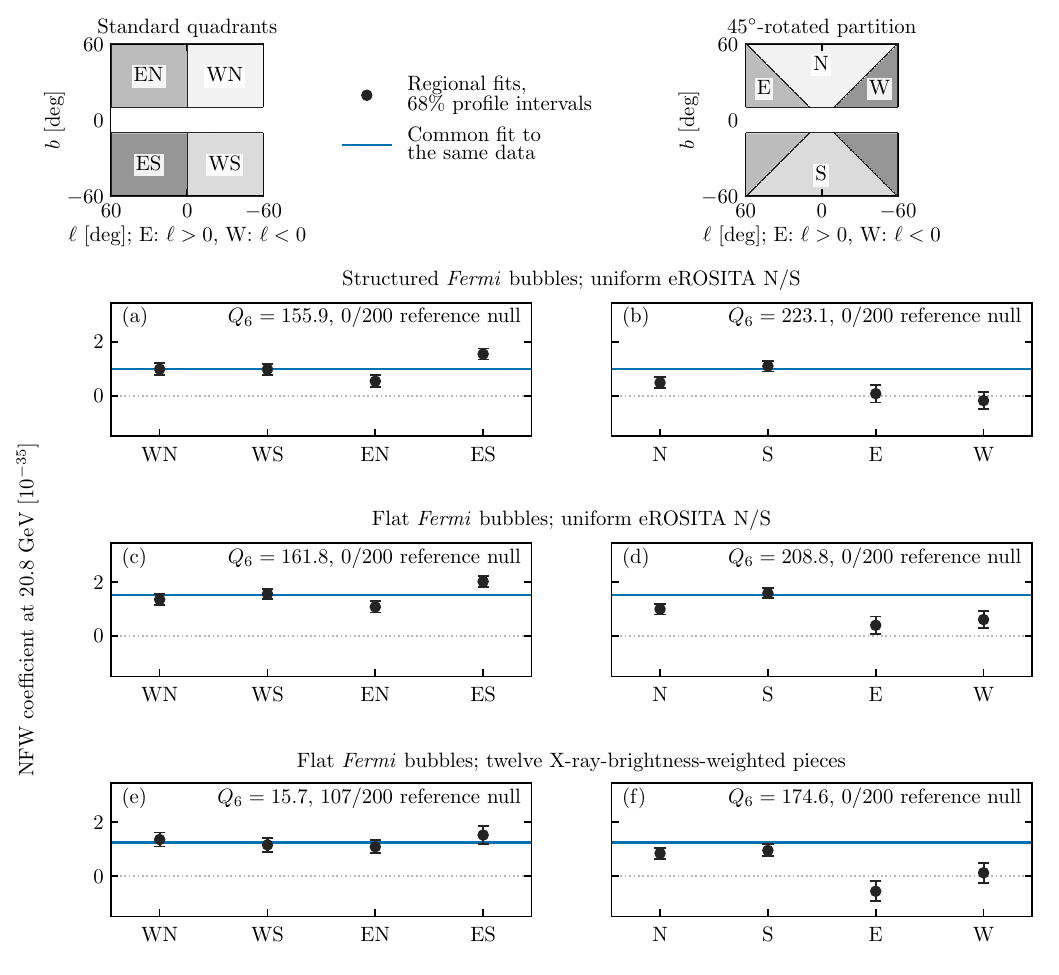}
\caption{Shown here are the NFW coefficients fitted independently in four
sky regions at 20.8\,GeV for Totani's baseline diffuse model in the
single-gas decomposition, with $2^\circ$ cells, for the standard
quadrants (left) and the partition rotated by $45^\circ$ (right), drawn
at top with the excluded plane in white; E and W denote $\ell>0$ and
$\ell<0$. The rows use, from top to bottom, the structured and flat
\fermi\ bubbles templates with uniform northern and southern eROSITA
bubbles templates, and the flat template with twelve
X-ray-brightness-weighted eROSITA bubbles pieces. Only with the twelve
pieces do the standard quadrants agree with a common normalization,
whereas the rotated partition disfavors one in every row. The filled
(black) circles are the regional coefficients with 68\%
profile-likelihood intervals, the solid (blue) line is the common
coefficient fitted to the same data, and the dotted (gray) line marks
zero. The plotted coefficients refer to the 20.8\,GeV
bin, whereas each displayed $Q_6$ sums the
regional-fit improvement over 5.6--130\,GeV.
Each panel also gives the number of 200 reference
null simulations exceeding its $Q_6$, with the
calibration scope described in
Sec.~\ref{sec:quadrantfinal}. Coefficients are
in units of
$10^{-35}\,\mathrm{cm^3\,s^{-1}\,GeV^{-2}\,MeV^{-1}\,sr^{-1}}$.}
\label{fig:regional_nfw_normalizations}
\end{figure*}

The quadrant tests ask whether the regional NFW amplitudes
agree, rather than whether they are individually positive.
Both the flat and structured \fermi\ bubbles fits favor
independent amplitudes
(Fig.~\ref{fig:regional_nfw_normalizations}, upper two
rows). For each fixed partition and for their maximum,
the observed statistics exceed all 200 reference simulations
at $2^\circ$ and all 50 at $1^\circ$.

These reference ensembles were generated from the
representative single-gas configuration,
\code{SOZ6R30T150C5}. They directly calibrate that
configuration and provide only an approximate comparison
for Totani's baseline and A2020. The directly calibrated
configuration exceeds all 200 simulations with either
\fermi\ bubbles template. Zero exceedances in 200
independent realizations imply a one-sided 95\% upper
bound of approximately 1.5\% on the corresponding tail
probability, not a measurement of a more extreme
significance.

Using twelve brightness-weighted eROSITA bubbles pieces brings the standard quadrants into agreement at 20.8\,GeV (Fig.~\ref{fig:regional_nfw_normalizations}e) and over the restricted energy range ($p=0.5$--0.9).
The partition rotated by $45^\circ$ still favors unequal amplitudes in every tested configuration, with no exceedances among 200 null simulations. 
In this partition, the polar wedges contain comparable positive contributions, while the east and west wedges nearer the plane contain little NFW-correlated emission (Fig.~\ref{fig:regional_nfw_normalizations}, right column).
The eROSITA pieces share boundaries with the standard quadrants, which may help explain why agreement is obtained for that partition alone.
Therefore, the restored north/south agreement does not extend to the full
NFW spatial pattern. 
For these foreground models, the regional fits disfavor assigning the entire excess to a single NFW normalization, although a spherical contribution can coexist with the remaining unmodeled structure.

We also replace the NFW template with annihilation maps from six HESTIA
Milky Way analogues \cite{2020MNRAS.498.2968L,Muru:2025vpz}, viewed
from 36 positions on the solar circle. Relative to NFW, their intrinsic
south/north $J$ ratios span 0.95--1.08 in the fitted region. Using these
maps leaves median fitted hemispheric amplitude ratios of 1.20--1.61
across five diffuse configurations, and every view fits
Totani's baseline worse than NFW at 20.8\,GeV, by 12--46 in
$-2\ln\mathcal L$. These six analogues therefore do not account for
the asymmetry in this foreground configuration. Other Galactic dark
matter distributions have not been tested. The map projection and
lobe-injection checks are described in Appendix~\ref{app:morphology}.

\subsection{A substructured \fermi\ bubbles model}
\label{sec:chz}

Cholis, Hu and Zhong \cite{Cholis:2026rba} describe the
\fermi\ bubbles with three components: flat bubbles with
updated boundaries, tilted prolate ``cocoons,'' and oblate
``inner bubbles.'' Their cocoon model has a southern-to-northern volume-emissivity ratio of ten. 
In their reported final high-latitude fit, the inner bubbles replace the halo-like template rather than being fitted alongside it.

We test whether an additional NFW component is preferred
when all three released templates are included. We hold
their spatial morphologies fixed, fold them through our
instrument response, and fit an independent normalization
for each component in every energy bin, together with
the accompanying foregrounds of Table~\ref{tab:res}.
This applies their released templates within our analysis;
it does not repeat their optimization of the bubbles
geometry.

Their model improves the fit substantially, by 480--560 in
$-2\ln\mathcal L$ over 5.6--130\,GeV relative to the flat template with
the NFW component included, mostly through the cocoons. However, an
NFW-correlated component remains on top of the full model in four of
the six representative configurations, with $Z_{2018}=6.9$--16.3 and
3070--5640 counts at 20.8\,GeV (Table~\ref{tab:chz}). In the two
five-ring configurations, it is weak, $Z_{2018}\le2.1$ at $2^\circ$ and
$\le3.2$ at $1^\circ$, and the enlarged flat bubbles alone already
remove most of it.
For Totani's baseline, the fitted NFW spectrum peaks near
21\,GeV; the full-range fit also gives negative amplitudes
below 3\,GeV. The preference over 5.6--130\,GeV remains
at $1^\circ$, with $Z_{2018}=8.3$--17.8 in the four
configurations with a strong additional component.
Replacing our Haslam template with their two Loop~I
shells gives $Z_{2018}=7.1$--18.7.

\begin{table*}[tbp]
\caption{
NFW fits with the three released substructured
\fermi\ bubbles templates of
Ref.~\cite{Cholis:2026rba} and the accompanying
foregrounds of Table~\ref{tab:res}. All fits
use $2^\circ$ cells and the PSF.
$A_{20.8}$ gives the full-region NFW counts
with 68\% profile intervals. $Z_{2018}$ is
evaluated over 5.6--130\,GeV, either in the
full region or outside both the 2014 mask
and the updated boundaries. The final column
is the full-region difference in
$-2\ln\mathcal L$ between the substructured
and flat \fermi\ bubbles fits, with NFW
included in both; negative values favor the
substructured templates. The additional NFW
preference is strong in four configurations
and weak in the two five-ring configurations.}
\label{tab:chz}
\centering
\fontsize{8.7}{10.2}\selectfont
\setlength{\tabcolsep}{3pt}
\begin{ruledtabular}
\begin{tabular}{lrrrr}
diffuse configuration & $A_{20.8}$ & $Z_{2018}$ & $Z_{2018}$, outside & $\Delta$\\
\hline
\code{SLZ6R30T150C2} 1-gas & 3071 [2453, 3690] & 6.9 & 1.2 & $-497$\\
\code{SOZ6R30T150C5} 1-gas & 4592 [4014, 5172] & 16.3 & 7.1 & $-483$\\
\code{SLZ6R30T150C2} 5-ring & 1590 [139, 3043] & 0.6 & 1.0 & $-525$\\
\code{SSZ10R30T150C5} 5-ring & 2356 [1353, 3481] & 2.1 & 2.4 & $-513$\\
Galp21 SA0\_R12 1-gas & 3832 [3226, 4440] & 10.4 & 6.2 & $-557$\\
A2020 & 5638 [5016, 6260] & 11.4 & 7.0 & $-506$\\
\end{tabular}
\end{ruledtabular}
\end{table*}

We test signal recovery on skies generated with the
released substructured templates and the representative
single-gas diffuse configuration. With the same spatial
templates held fixed in the fit, an injected halo with
Totani's spectrum is recovered at 0.95--1.03 of its input
value in each energy bin from 4.3\,GeV upward. Fifty
realizations without a halo give
$2\,[-460,670]$ fitted NFW counts at 20.8\,GeV.

Fitting those same no-halo skies with the flat \fermi\
bubbles template instead produces
$3300\,[2870,3870]$ spurious NFW counts, with
$Z_{2018}\approx17.5$. The recovered spectrum is broad
and resembles the component fitted to the data with
the flat template in this configuration, which contains
6150 counts. Unmodeled bubbles substructure can therefore
produce a substantial apparent NFW contribution. The
matched-template test establishes recovery for the
specified injection and foreground; it does not test
signal absorption during optimization of the bubbles
morphology.

We repeat the quadrant tests for Totani's baseline, the
representative single-gas configuration, and A2020.
With the substructured \fermi\ bubbles templates, the
six-bin statistics fall from 104--162 to 17--31 for the
standard partition and from 114--209 to 50--86 for the
rotated partition. We calibrate each configuration with
200 null simulations drawn from its own
common-normalization fit. In the order listed above,
3, 12, and 93 simulations exceed the observed
standard-partition statistics, while 0, 0, and 1 exceed
the rotated-partition statistics. The rotated values
remain 53--92 at $1^\circ$ and 40--74 when their
Loop~I shells replace the Haslam template.

Outside both the 2014 mask and the updated boundaries,
the nominal NFW preference spans $Z_{2018}=1.0$--7.1
at $2^\circ$ and 1.1--7.2 at $1^\circ$, exceeding
$6\sigma$ in three of the six configurations. The
substructured templates thus account for a substantial
part of the emission and reduce the regional
discrepancies. An additional component is nevertheless
preferred in some foreground configurations, and the
rotated partition continues to disfavor a common NFW
normalization in the three configurations tested.

With the substructured \fermi\ bubbles templates, the
420\,GeV $W^+W^-$ spectrum gives $p\ge0.05$ in four
of the six configurations. It is disfavored in the
representative single-gas configuration and A2020,
with $p=0.027$ and $1.6\times10^{-4}$, respectively.
The four acceptable cases include the two five-ring
configurations in which the additional NFW component
is weak, so spectral agreement alone does not establish
a signal. Where the component is strongly preferred
and the benchmark shape is acceptable, the fitted
coefficient is $(2.4$--$3.6)\times10^{-25}\,\mathrm{cm^3\,s^{-1}}$,
below the values obtained with the flat and structured
\fermi\ bubbles templates.

\subsection{Signal absorption and foreground-induced excesses}
\label{sec:injection}
\label{sec:simresults}

Rebuilding Totani's residual \fermi\ bubbles templates after an
injection exposes signal losses that are absent when the templates
are held fixed. For the empirical Totani spectrum, we recover 93\%
of the injected contribution at 20.8\,GeV, but only 42\% at the
4.3\,GeV construction energy. An injection with a constant template
coefficient even gives negative recovery slopes at 1.5 and 2.6\,GeV
despite adding positive emission.
After numerical refinement, the fixed-template controls
recover 0.99--1.01 of the injected contribution
(Table~\ref{tab:injection}). The losses that remain
when the templates are rebuilt therefore arise from
the construction procedure, rather than insufficient
numerical accuracy in the fits.

The signal absorption is energy dependent and changes the recovered
spectrum. The constant coefficient used in the test does not correspond
to a flat $E^2dN/dE$ spectrum, and the two injected spectra do not
establish that an arbitrary signal would acquire a 20\,GeV peak.
For the empirical Totani injection, most of the signal survives near
the peak, while the lower-energy spectrum is more strongly affected
by leakage into the residual \fermi\ bubbles templates.

Simulations without an injected halo can also return a large positive
NFW component. In the final ensembles, skies generated with the
structured \fermi\ bubbles template and fitted with the flat template
give approximately 4700 spurious counts and
$Q^+=1750\,[1655,1836]$ over nine energy bins. Totani's residual
templates return approximately 3300--4300 counts and $Q^+=870$--1440
on skies without a halo, including those generated with the flat
\fermi\ bubbles template. These spurious spectra resemble the
\fermi\ bubbles spectrum, rather than the narrower peak found in
the data with the structured template (Fig.~\ref{fig:specsims}).
Foreground mismodeling can therefore produce a large nominal template
significance, although these simulations do not reproduce all the
observed spectral properties.

When the injected and fitted annihilation spectra match,
the twenty-sky test recovers
$\sv=4.9\,[4.6,5.2]\times10^{-25}\,\mathrm{cm^3\,s^{-1}}$
for a 500\,GeV $b\bar b$ injection at
$5\times10^{-25}\,\mathrm{cm^3\,s^{-1}}$. The recovered
value changes by less than 2\% across the tested
lower-energy cutoffs. By contrast, fitting the empirical
Totani spectrum with a different spectral family gives
a smaller coefficient. That difference reflects the
imposed spectral shape, not a failure to recover the
matching annihilation injection.

\section{Discussion and conclusions}
\label{sec:discussion}
\label{sec:conclusions}

Our reanalysis reproduces Totani's excess under the original assumptions, and over 5.6--130\,GeV, the fits with the instrument response included retain NFW-correlated emission with both the flat and structured \fermi\ bubbles templates. 
The small and insignificant NFW amplitude obtained with the structured template at the original $10^\circ$ binning therefore does not establish an absence of this component, nor does a background-only preference for the structured template, since with the NFW component included the flat template fits better in every representative configuration (Sec.~\ref{sec:twodeg}).
However, the regional normalizations disagree, when annihilation in a smooth halo predicts a common normalization (Secs.~\ref{sec:erositafreedom} and \ref{sec:quadrantfinal}). 
Therefore, the finer-binning results disfavor assigning the full signal to annihilation in the adopted smooth NFW halo. 
In addition, the inferred spectrum depends on the foreground model (Sec.~\ref{sec:dm}), and the amplitude changes by a factor of 1.4--2.1 with the \fermi\ bubbles template alone (Sec.~\ref{sec:twodeg}), so that the
annihilation cross section, and its consistency with a specific particle model, cannot be determined independently of the foreground description.

The fits outside the 2014 \fermi\ bubbles mask are less
sensitive to the brightness pattern adopted inside that
mask, but not to its boundary. Excluding cells touched
by the updated boundaries lowers the nominal preference
from approximately $4$--$16\sigma$ to $1$--$7\sigma$.
The outside-region amplitude also varies by about a
factor of two across diffuse configurations. These
results do not establish a uniformly significant
component beyond the bubbles. They show that the
inference there remains sensitive to both the diffuse
foreground and the boundary definition.

The injection and null simulations help distinguish sensitivity from
foreground bias. Fits with matching templates recover an injected
signal, but an incorrect \fermi\ bubbles template can produce an
NFW component of comparable amplitude even when none was injected.
Rebuilding Totani's residual templates after injection also transfers
some of the signal into the fitted foreground and changes the recovered
spectrum. These effects limit what can be inferred from a large
template significance alone. At the same time, the simulated false
components do not reproduce all the spectral properties of the data
(Sec.~\ref{sec:injection}), so the tests do not establish that foreground
mismodeling explains the entire excess.

The eROSITA bubbles tests show how foreground structure can affect the
inferred morphology. Dividing the brightness-weighted template into
independent regions restores north/south agreement, and injections show
that fitting a spherical signal with an inadequate eROSITA foreground
can produce an apparent asymmetry. However, uniform eROSITA bubbles
templates with the same spatial freedom fit at least as well once the
NFW component is included. The X-ray structure that reduces the
asymmetry also lies in the reconstructed eastern map. Its ability to
restore hemispheric agreement is therefore insufficient to identify
it as the gamma-ray counterpart of the X-ray emission.

The substructured \fermi\ bubbles templates substantially
improve the fit and weaken the preference for additional
NFW-correlated emission, especially in the two five-ring
configurations. Our results therefore support an important
role for bubbles substructure. With the released
morphologies held fixed, however, an additional component
remains strongly preferred in four of the six
representative configurations. The rotated partition
also disfavors a common normalization in the three
configurations subjected to that test. Whether
additional emission is required thus depends on both
the diffuse foreground and the description of the
bubbles; its persistence in some fits does not identify
it as dark matter.

Even after north/south agreement is restored, other regions require
different NFW normalizations. The partition rotated by $45^\circ$
still favors independent amplitudes, and the six HESTIA analogues do
not account for the asymmetry in the foreground configuration examined
(Sec.~\ref{sec:quadrantfinal}). This is evidence against describing
the full excess with the adopted NFW pattern and a single coefficient,
not simply uncertainty in that coefficient. 
The test does, however, depend on the accompanying foreground model. In simulated skies that contain a spherical NFW signal and eROSITA bubbles emission distributed as the twelve brightness-weighted pieces, fitting two uniform eROSITA bubbles templates yields a south/north ratio of 1.35, close to the 1.37 found in the data, whereas matching templates recover 1.05 (Appendix~\ref{app:morphology}). 
Therefore, foreground mismatch alone can make a spherical signal appear to require different regional normalizations.

These findings weaken the gamma-ray support for a common dark matter
origin with the LZ event. 
In particular, where the 420\,GeV $W^+W^-$ spectrum fits acceptably, its fitted coefficient is $(5.0$--$8.4)\times10^{-25}\,\mathrm{cm^3\,s^{-1}}$ for our halo normalization, a factor of 3--5 below the Sommerfeld-enhanced halo rate of $2.7\times10^{-24}\,\mathrm{cm^3\,s^{-1}}$ in the proposed scenario \cite{Yamashita:2026ump}. With the substructured \fermi\ bubbles model, it is lower still, $(2.4$--$3.6)\times10^{-25}\,\mathrm{cm^3\,s^{-1}}$ (Sec.~\ref{sec:chz}).
A successful fit
to a freely normalized spectrum in some configurations neither removes
the spatial discrepancy nor gives a stable determination of the
particle properties. We have not fitted the LZ likelihood
or tested the complete particle model. Our conclusions concern the
astrophysical evidence used to motivate the connection, not the
significance of the recoil event reported by LZ \cite{LZ:2026axp}.

We conclude that, for the foreground models examined here, the spatial and spectral results disfavor interpreting the full gamma-ray excess as annihilation in the adopted smooth NFW halo, and therefore weaken the present astrophysical case for a common dark matter origin of the gamma-ray excess and the LZ event. 
They do not, however, exclude a dark matter contribution. 
A halo-like component persists outside the 2014 \fermi\ bubbles mask in every foreground model tested (Sec.~\ref{sec:outside}), although with the flat template its significance falls to 1--7$\sigma$ when the enlarged boundaries of Ref.~\cite{Cholis:2026rba} are also excluded (Sec.~\ref{sec:chz}), and our foreground tests do not explain all of the residual emission. 
The difficulty lies in its entanglement with the large Galactic structures, the \fermi\ bubbles, Loop~I, and the eROSITA bubbles, whose morphologies are not yet independently determined. 
Independent measurements of these foregrounds, including native eastern eROSITA data, entering fits that allow both foreground emission and a halo signal, would help
separate these contributions. 
It remains to be determined whether a dark matter component of the halo-like residual, once disentangled from these structures, is consistent with the mass scale indicated independently by the LZ event.

\section*{Data availability}
The LAT photon and spacecraft data, response functions, and 4FGL-DR4
catalog are available from the Fermi Science Support Center. The 2012
GALPROP grid accompanies Ref.~\cite{Ackermann:2012pya}. 
The Galp21 configurations were computed with GALPROP v57 (\url{https://galprop.stanford.edu/}) \cite{Porter:2021tlr}. Their gamma-ray templates on our analysis grid and the native GALPROP outputs are available from the authors on request.
The \fermi\ bubbles, Loop~I, and eROSITA bubbles templates are based on
the sources described above. The authors of Ref.~\cite{Muru:2025vpz}
supplied the HESTIA particle data and halo profiles, which are available
from them on reasonable request. The analysis code, derived templates, simulated outputs, and tables are released at \url{https://github.com/kabazajian/totani}, with Zenodo
archives of the software (\url{https://doi.org/10.5281/zenodo.23420779}) and
of the data (\url{https://doi.org/10.5281/zenodo.23420896}).

\begin{acknowledgments}
We thank Eiichiro Komatsu and Shin'ichiro Ando for stimulating discussions, and M.~M.~Muru and collaborators for sharing the HESTIA particle data and halo profiles. KNA is partially supported by the U.S. National Science Foundation Theoretical Physics Program under Grant No.~PHY-2609954. OM is supported by the U.S. National Science Foundation under Grant No.~2418730. Claude Code (Anthropic) and ChatGPT (OpenAI) were used for software implementation, numerical execution, initial drafting of some text, and draft proofreading.  The authors are responsible for the analysis and its interpretation.
\end{acknowledgments}

\appendix
\section{Data reduction and response checks}
\label{app:response}

We select events with \code{evclass=512} and \code{evtype=3}, a
$100^\circ$ zenith cut, and an $86^\circ$ extraction radius centered
on the Galactic Center. The initial extraction spans
1.16\,GeV--1.06\,TeV; the fits use its nine bins below 130\,GeV.
Applying \code{gtmktime} with
\code{DATA\_QUAL==1 \&\& LAT\_CONFIG==1 \&\& ABS(ROCK\_ANGLE)<52}
and no ROI cut retains $1.58\times10^7$ events from 2008 August 4 to
2023 August 3. The count maps have $960\times960$ pixels. We calculate
the livetime cube with $\Delta\cos\theta=0.025$ and $1^\circ$ pixels
and tabulate exposure at fourteen extraction-bin edges.

For energy-independent spatial templates, we multiply the intensity
by pixel solid angle and $\int_e A(E)\,dE$. The exposure is interpolated
log-linearly and integrated on a 32-point grid, assuming a flat
differential spectrum within each energy bin. GALPROP spectra are
integrated piecewise and multiplied by the exposure at the geometric-mean
energy. Of 2651 catalog sources in the full square, 1425 lie at
$|b|\ge10^\circ$. Extended-source masks remove 240 sources, leaving
2411 in the source template. The high-latitude masks cover the Small
Magellanic Cloud, Centaurus~A lobes, FHES J1626.9$-$2431, and FHES
J1723.5$-$0501.

We evaluate \code{gtpsf} toward the Galactic Center using the selection's
livetime cube, average the response with the exposure and an
$E^{-2.4}$ spectrum, and fit it with three Gaussians. Longitude widths
are scaled by $1/\cos b$, and source kernels are centered at their
catalog positions. The kernels conserve flux and reproduce the 95\%
containment radius to within 5\% in every energy bin and the 68\%
radius to within 4\% over 1.5--4.3\,GeV. At 7.3 and 12.3\,GeV, the
kernel 68\% radius is 13\% smaller than tabulated. From 20.8\,GeV
upward, the 68\% radius is smaller than the $0.125^\circ$ pixel size.

We benchmark against \code{gtsrcmaps} in three
$20^\circ\times20^\circ$ fields centered at
$(\ell,b)=(0^\circ,16^\circ)$, $(0^\circ,40^\circ)$, and
$(-40^\circ,-40^\circ)$, with identical event class, livetime,
and pixelization. Above 5.6\,GeV, cell-integrated NFW counts agree
to 0.1\% and structured \fermi\ bubbles counts to 1--2.4\%. At most
3\% of any of nine bright sources' counts fall in different cells
than in Fermitools (4.7\% over the wider tested range). Extending the
templates and exposure to $\pm62^\circ$ changes the 20.8\,GeV
amplitude by at most 0.1\%, $Q^+$ by 0.4\%, and the restricted-range
flat-versus-structured likelihood difference by one unit at either
binning.

Including the PSF matters even when its containment radius is small
relative to a cell. At $2^\circ$, it changes the 20.8\,GeV amplitude
by $+3$ to $+7$\% with the structured \fermi\ bubbles template,
$-12$ to $+1$\% with the flat template, and $-2$ to $-13$\% with
Totani's residual templates. The changes span $-16$ to $+15$\% at
$1^\circ$ and reach tens of percent in the lowest energy bins.
The restricted-range comparison between the flat and structured
templates shifts by 420--430 likelihood units. Point sources show
the largest redistribution between cells at 100\,GeV (14\% with
$1^\circ$ cells and 8\% with $2^\circ$ cells, compared with at most
0.2\% for the diffuse and \fermi\ bubbles templates),
which motivated the comparison at the exact source positions.

\section{Diffuse-model construction}
\label{app:diffuse}

\subsection{The 2012 grid}
\label{app:galprop}

The 2012 grid varies four cosmic-ray source distributions, propagation
halo heights of 4, 6, 8, and 10\,kpc, halo radii of 20 and 30\,kpc,
spin temperatures of 150\,K and $10^5$\,K, and $E(B-V)$ cuts of 2 and
5 magnitudes \cite{Ackermann:2012pya}. The released HEALPix cubes
contain 38 energies. We integrate each component over the analysis
bins using power laws between tabulated energies. The five gas groups
combine native annuli 1, 2--5, 6--12, 13, and 14--17. The three inverse
Compton components use optical, infrared, and microwave target photons.

The 51 bremsstrahlung ring maps sum to the released total, including
its ionized-gas contribution, which is not added separately. The three
inverse Compton maps also sum to their released total. Toward the
Galactic Center, the pion-decay $E^2dN/dE$ spectrum peaks near
600\,MeV and has a differential slope of $-2.64$ over 1--10\,GeV.
Independently generated Galp21 gas templates agree with the 2012
baseline to factors of 0.98--1.02 at 1.5\,GeV. These checks test units
and component accounting; they do not validate the full foreground
model against the data.

\subsection{Galp21 realizations}
\label{app:galp21}

The Galp21 models use pure-disk, equal disk/spiral, or pure-spiral source
distributions with the F98 or R12 radiation field. Gas emissivities
follow the three-dimensional model of Ref.~\cite{Johannesson:2018bit}
and are rescaled to observed, dust-corrected H\,\textsc{i} and CO
columns. We use 210 of the 233 cubes delivered for each model. They
contain three gas phases in seventeen rings and inverse Compton
emission from three radiation-field components, calculated in both
the isotropic and anisotropic treatments. The
$3600\times1200$ grid has $0.1^\circ$ pixels, covers $|b|\le60^\circ$,
and contains 44 energies from 100\,MeV to 2\,TeV.

We obtain the longitude orientation and wrapping from the FITS headers.
Differential intensities, in
$\mathrm{cm^{-2}\,s^{-1}\,sr^{-1}\,MeV^{-1}}$, are sampled at the
nearest native pixel and integrated with the same log-log scheme used
for the 2012 maps. The ring-group boundaries are 1.485, 3.500, 7.991,
and 10.0\,kpc.
The ring inverse Compton cubes sum to the delivered totals to $10^{-10}$. The maps were generated with GALPROP v57.0.3032 after four corrections: the CO ring cube's reference pixel was ignored when it was read (removing H2 from all non-HEALPix output), the output pixel size was rounded to a power of two, the bremsstrahlung ring renormalization indexed the gas arrays out of bounds, and a 32-bit counter overflowed in the per-ring map writers. Negative pixels in the input CO ring cube (0.05\% of pixels) were set to zero
before the runs, and the resulting gas maps contain no negative pixels.

The main models use the release example's radial prescription
\begin{equation}
 X_{\rm CO}=10^{20}\,10^{-0.4+0.066R/\mathrm{kpc}}\,
 \mathrm{cm^{-2}\,(K\,km\,s^{-1})^{-1}},
\end{equation}
which is held constant beyond 15\,kpc.

At $10^\circ$ binning, degrading the Galp21 maps to $N_{\rm side}=128$
or smoothing to $0.46^\circ$ changes the background-only
$-2\ln\mathcal L$ by $-17$ to $+8$. Replacing individual components
indicates that the inverse Compton map accounts for most of the larger
Galp21 fit deficit. Relative to the best 2012 model, 2012 gas with
Galp21 anisotropic inverse Compton emission gives a deficit of
9800--11,300, compared with 1400--1800 for Galp21 gas with 2012 inverse
Compton emission and 2600--4500 for Galp21 isotropic inverse Compton
emission.
These comparisons identify the inverse Compton
maps as the main source of the poorer fits
for the particular Galp21 realizations tested.
They do not establish a physical preference
against anisotropic scattering.

\section{Comparisons with the original spatial binning}
\label{app:coarse}
\label{sec:repro}
\label{sec:figure}
\label{sec:conditional}
\label{sec:selresults}

This appendix uses $10^\circ$ cells and nine energy bins unless
stated otherwise. Totani's baseline gives $6412\,[5904,6898]$ counts
at 20.8\,GeV and $\mathrm{TS}=172$. The positive quadrant amplitudes
sum to 0.93 of the joint amplitude. The reconstructed images recover
the negative northern band at $b=10^\circ$--$20^\circ$. Substituting
the flat or structured \fermi\ bubbles template gives 5006 and
1575 counts, respectively. These tests reproduce the original
procedure; the final results use finer bins.

With the accompanying foreground templates held fixed, the original
diagnostic is passed by 236, 257, and 132 of the 257 configurations
using the residual, flat, and structured \fermi\ bubbles templates,
respectively. Passing requires $\mathrm{TS}_{20.8}>4$ and positive
amplitudes in all four quadrants. It is not a global dark matter test.
The LAT model containing residual-derived emission is excluded from
this ranking.

\begin{figure*}[tp]
\centering
\includegraphics[width=\textwidth]{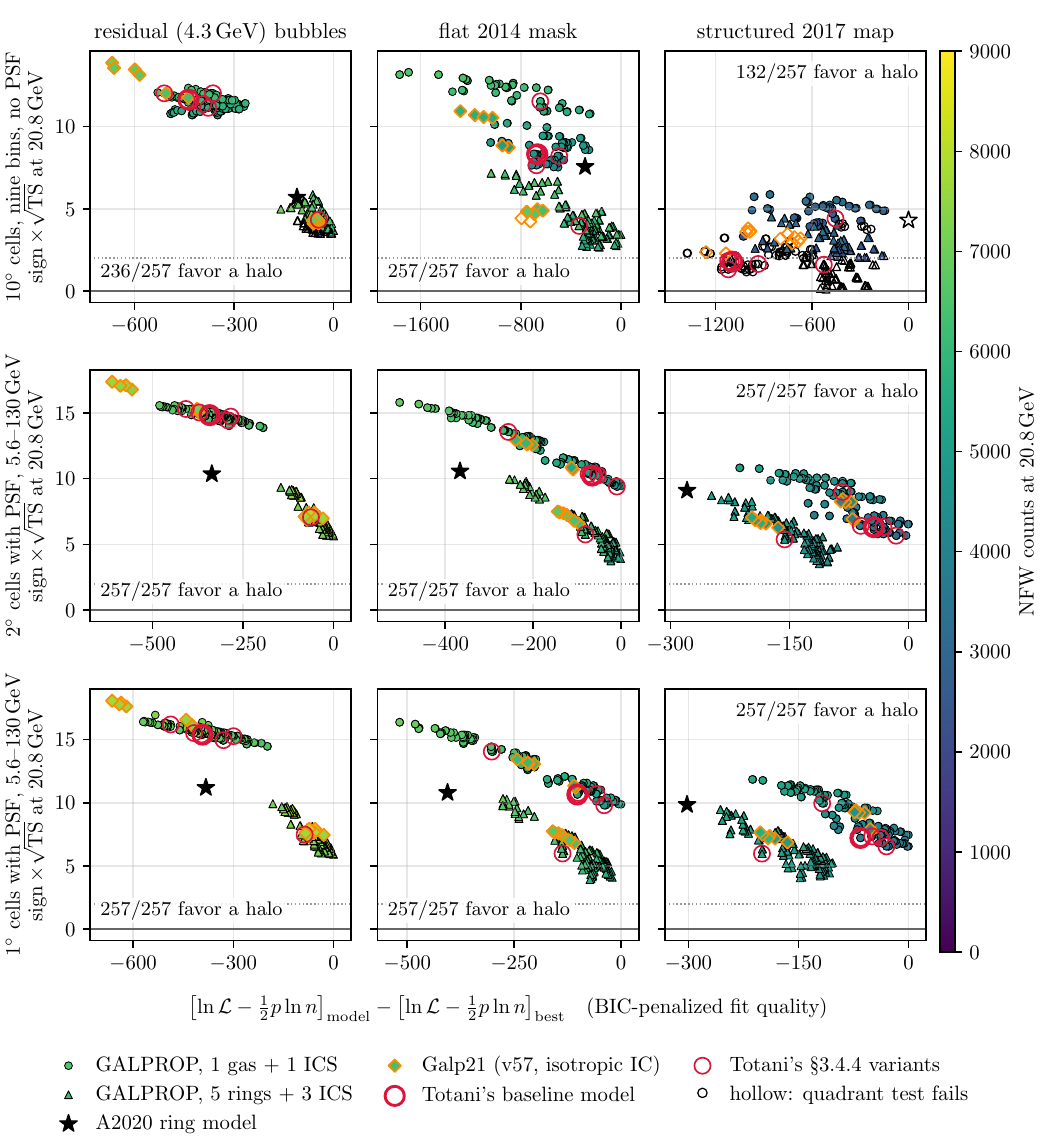}
\caption{Shown here is the signed 20.8\,GeV significance against
BIC-penalized background fit quality for the diffuse configurations,
for the three \fermi\ bubbles descriptions (columns) at the original
$10^\circ$ binning over nine bins (top) and at $2^\circ$ and $1^\circ$
with the response included over 5.6--130\,GeV (middle, bottom). Each
point is a diffuse configuration; color gives NFW counts and hollow
markers fail quadrant positivity. Accompanying templates are Loop~I
shells, uniform north/south eROSITA bubbles, isotropic emission, and
catalog sources. Rings identify Totani's configurations, the star is
A2020, and diamonds are Galp21 models with isotropic inverse Compton
emission. ``Favor a halo'' refers to the original diagnostic. The
structured-template amplitudes, which fail the diagnostic in about
half of the configurations at $10^\circ$, pass it in every
configuration at the finer binnings.}
\label{fig:ranking}
\end{figure*}

The top row of Fig.~\ref{fig:ranking} shows these fits. Each point
is one diffuse configuration, the vertical axis is the signed
20.8\,GeV significance, and the horizontal axis is the BIC-penalized
background-only $-2\ln\mathcal L$ relative to the best configuration
in the panel. At this binning, the fitted 20.8\,GeV amplitude spans
5500--8500 counts with the residual templates, 4300--7900 with the
flat template, and 170--4600 with the structured template. The
\fermi\ bubbles template therefore sets the amplitude, and the diffuse
model moves it within each panel.

The lower rows repeat the comparison at $2^\circ$ and $1^\circ$ with
the response included, over 5.6--130\,GeV, with the same companions.
All 257 configurations favor a halo under every template at both
binnings. The 20.8\,GeV amplitudes are 6100--8700 (residual),
4700--7900 (flat), and 3100--6300 counts (structured) at $2^\circ$,
and 6400--8900, 4900--8100, and 3500--6700 at $1^\circ$. The
dependence of the significance on fit quality within each
decomposition persists at the finer binnings, including for the
structured template.

The relation between nominal NFW significance
and foreground fit quality depends on the
template choice and decomposition. For the
residual \fermi\ bubbles templates, the
single-gas/five-ring Spearman coefficients
are $+0.02/-0.08$ without a halo,
$-0.35/-0.38$ with an injected signal,
and $-0.05/-0.53$ in the data. With the
structured template, they are
$+0.49/+0.04$, $+0.39/-0.30$, and
$+0.64/-0.41$, respectively. Correlations,
including anticorrelations, can therefore
occur when a signal is present. Moreover,
the two diagnostics share the
background-only likelihood, so their
correlation is not an independent test
of foreground contamination.

Tables~\ref{tab:selection} and \ref{tab:galp21} give the coarse
selections. Each decomposition includes A2020 among its 129 entries,
leaving 257 distinct configurations overall. Without Totani's
residual templates, single-gas fits select the structured \fermi\
bubbles template and five-ring fits select the flat template. The
resulting amplitude and signed-coefficient differences are specific
to this binning.

\begin{table*}[t]
\caption{Coarse selection with $10^\circ$ cells and nine energy bins. ``Pass''
requires $\mathrm{TS}_{20.8}>4$ and four positive quadrant amplitudes.
Each selection contains 128 models and A2020. Parentheses give
percentages clearing the BIC threshold and winning the comparison.
NFW counts are model ranges. $\sv$ is for $b\bar b$ at 500\,GeV,
in $10^{-25}\,\mathrm{cm^3\,s^{-1}}$, with bracketed model ranges.
For the second row, the final percentages
give the fractions of positive fitted
coefficients in fit-quality terciles,
ordered from best to worst.}
\label{tab:selection}
\centering
\fontsize{8}{9.5}\selectfont
\setlength{\tabcolsep}{3pt}
\begin{ruledtabular}
\begin{tabular}{llp{0.30\textwidth}ccc}
gas & construction & selected (clears / wins, \%) & pass & counts & $\sv$ \\
\hline
1-gas & allowed & residual (100/100) $\to$ shells $+$ NPS (100/89) $\to$ eROSITA bright.\ N/S (99/99) & 129/129 & 5456--8437 & $+4.0$ [2.8, 7.7]; 100\% \\
1-gas & excluded & structured (100/57) $\to$ eROSITA unif.\ N/S (100/100) $\to$ Haslam $\to$ solar IC & 0/129 & 506--3487 & $-2.2$ [$-5.6$, 3.9]; 23/5/0\% \\
5-ring & allowed & residual (100/100) $\to$ shells $+$ NPS (100/98) $\to$ solar IC (99/100) & 43/129 & 6187--8549 & $+5.9$ [4.8, 8.1]; 100\% \\
5-ring & excluded & flat (100/64) $\to$ eROSITA unif.\ outer N/S (100/51) $\to$ Haslam $\to$ solar IC & 129/129 & 5456--7424 & $+5.1$ [1.2, 8.9]; 96\% \\
\end{tabular}
\end{ruledtabular}
\end{table*}
\begin{table*}[t]
\caption{Galp21 fits with $10^\circ$ cells and nine bins. Rows summarize six
models per inverse Compton treatment and diffuse decomposition.
Imposed selections follow Table~\ref{tab:selection}; final columns
use individual selections excluding residual \fermi\ bubbles
templates. NFW counts are ranges, ``pass'' uses the coarse positivity
diagnostic, and $\sv$ is a median in
$10^{-25}\,\mathrm{cm^3\,s^{-1}}$ for $b\bar b$ at 500\,GeV.
Rank compares the diffuse-only fit with the 129-entry 2012 ensemble.}
\label{tab:galp21}
\centering
\fontsize{8}{9.5}\selectfont
\setlength{\tabcolsep}{3pt}
\begin{ruledtabular}
\begin{tabular}{llcccccccc}
IC & gas & rank & \multicolumn{2}{c}{construction selection} & \multicolumn{2}{c}{2012 no-construction selection} & \multicolumn{3}{c}{own selection}\\
 & & & counts; pass & $\sv$ & counts; pass & $\sv$ & bubbles & pass & $\sv$ \\
\hline
aniso.\ & 1-gas & 108--122 & 5906--6468; 6/6 & $+5.8$ & 1631--2137; 0/6 & $-0.7$ & structured 6/6 & 0/6 & $-0.8$ \\
aniso.\ & 5-ring & 98--126 & 6580--7087; 1/6 & $+6.2$ & 6637--6974; 3/6 & $+6.6$ & structured 6/6 & 0/6 & $+1.4$ \\
iso.\ & 1-gas & 42--61 & 6131--6749; 6/6 & $+5.7$ & 1181--1744; 0/6 & $-1.6$ & structured 4, flat 2 & 2/6 & $-1.3$ \\
iso.\ & 5-ring & 56--96 & 6805--7255; 3/6 & $+6.1$ & 6131--6524; 3/6 & $+5.8$ & structured 6/6 & 0/6 & $+0.5$ \\
\end{tabular}
\end{ruledtabular}
\end{table*}

With NFW included, flat \fermi\ bubbles fit better than structured
\fermi\ bubbles in every listed single-gas configuration, despite
the background-only selection of the structured template
(Table~\ref{tab:conditional}). Removing eROSITA templates worsens
$-2\ln\mathcal L$ by about 1600--3000 without residual templates.
With them, the eROSITA improvement falls below the eighteen-parameter
BIC cost. Residual templates also outperform flat and structured
templates, although they are constructed from the fitted data.

\begin{table*}[t]
\caption{Conditional $10^\circ$, nine-bin fits. ``With'' and ``without''
refer to selections allowing or excluding residual \fermi\ bubbles
templates. $H_1$ adds signed NFW amplitudes to $H_0$, refitting all
nuisance parameters. TS is the nine-bin sum and 20.8\,GeV value;
$A$ gives 20.8\,GeV counts with its 68\% interval. $\sv$ is the
500\,GeV $b\bar b$ coefficient in
$10^{-25}\,\mathrm{cm^3\,s^{-1}}$, with interval and TS against
zero. Final columns give changes in $-2\ln\mathcal L(H_1)$ when
substituting \fermi\ bubbles templates or adding/removing the
eROSITA pair. Negative changes indicate improved fits. Galp21 rows
use the isotropic inverse Compton maps. These are
not the final response-folded results.}
\label{tab:conditional}
\centering
\fontsize{7.8}{9.3}\selectfont
\setlength{\tabcolsep}{3pt}
\begin{ruledtabular}
\begin{tabular}{llccccc}
configuration & model & TS sum / 20.8 & $A$ [68\%] & $\sv$ [68\%] (TS) & flat / struct. / constr. & eROSITA \\
\hline
\code{SLZ6R30T150C2} 1-gas & with & 682\,/\,102 & 5962 [5347, 6538] & $+3.1$ [2.9, 3.3] (168) & $+6905$\,/\,$+8369$\,/\,0 & remove $+41$ \\
 & without & 957\,/\,1.3 & 787 [106, 1487] & $-4.7$ [$-5.0$, $-4.4$] (267) & $-543$\,/\,0\,/\,$-4335$ & remove $+2085$ \\
\code{SLZ6R30T150C2} 5-ring & with & 88\,/\,28 & 6580 [5661, 7488] & $+5.5$ [5.0, 6.0] (75) & $+7264$\,/\,$+8636$\,/\,0 & add $-33$ \\
 & without & 67\,/\,15 & 6243 [5037, 6937] & $+5.6$ [5.1, 6.1] (60) & 0\,/\,$+1388$\,/\,$-3383$ & remove $+2904$ \\
\code{SLZ10R30T150C5} 1-gas & with & 789\,/\,105 & 6018 [5448, 6643] & $+2.8$ [2.5, 3.0] (132) & $+6449$\,/\,$+7755$\,/\,0 & remove $+35$ \\
 & without & 1219\,/\,0.5 & 506 [$-207$, 1196] & $-5.6$ [$-5.9$, $-5.3$] (364) & $-437$\,/\,0\,/\,$-4041$ & remove $+2082$ \\
\code{SLZ10R30T150C5} 5-ring & with & 78\,/\,21 & 6524 [5614, 7484] & $+5.2$ [4.6, 5.8] (44) & $+6523$\,/\,$+7731$\,/\,0 & add $-37$ \\
 & without & 30\,/\,8.2 & 5906 [4171, 6783] & $+2.2$ [1.2, 3.3] (5) & 0\,/\,$+1281$\,/\,$-3051$ & remove $+2977$ \\
\code{SOZ6R30T150C5} 1-gas & with & 501\,/\,83 & 5512 [4906, 6134] & $+5.4$ [5.1, 5.6] (468) & $+5770$\,/\,$+7125$\,/\,0 & remove $+46$ \\
 & without & 68\,/\,13 & 2306 [1674, 2973] & $+0.3$ [0.1, 0.6] (2) & $-424$\,/\,0\,/\,$-3472$ & remove $+2345$ \\
\code{SSZ10R30T150C5} 5-ring & with & 92\,/\,25 & 6524 [5613, 7486] & $+5.5$ [5.0, 6.0] (64) & $+6396$\,/\,$+7604$\,/\,0 & add $-32$ \\
 & without & 33\,/\,11 & 5962 [4423, 6726] & $+3.2$ [2.3, 4.0] (12) & 0\,/\,$+1295$\,/\,$-2873$ & remove $+2994$ \\
A2020 & with & 141\,/\,31 & 8437 [7438, 9072] & $+7.7$ [7.1, 8.3] (128) & $+4782$\,/\,$+5671$\,/\,0 & remove $+27$ \\
 & without & 124\,/\,10 & 3487 [2457, 4608] & $+3.9$ [3.4, 4.3] (59) & $-361$\,/\,0\,/\,$-3101$ & remove $+1677$ \\
Galp21 SA0\_R12 1-gas & with & 651\,/\,110 & 6131 [5557, 6753] & $+4.0$ [3.7, 4.2] (278) & $+7112$\,/\,$+8397$\,/\,0 & remove $+56$ \\
 & without & 505\,/\,2.9 & 1181 [492, 1898] & $-3.3$ [$-3.6$, $-3.0$] (127) & $-450$\,/\,0\,/\,$-4482$ & remove $+1974$ \\
Galp21 SA0\_R12 5-ring & with & 88\,/\,27 & 7087 [6105, 8368] & $+6.1$ [5.6, 6.6] (75) & $+7619$\,/\,$+8794$\,/\,0 & add $-18$ \\
 & without & 116\,/\,20 & 6524 [5466, 7225] & $+6.1$ [5.5, 6.5] (96) & 0\,/\,$+1101$\,/\,$-3746$ & remove $+2506$ \\
Galp21 SA100\_F98 1-gas & with & 840\,/\,140 & 6749 [6172, 7331] & $+6.4$ [6.2, 6.6] (760) & $+7264$\,/\,$+8338$\,/\,0 & remove $+41$ \\
 & without & 212\,/\,6.3 & 1687 [1015, 2357] & $-1.5$ [$-1.8$, $-1.2$] (28) & $-381$\,/\,0\,/\,$-4143$ & remove $+1600$ \\
Galp21 SA100\_F98 5-ring & with & 97\,/\,25 & 7030 [6053, 7979] & $+6.0$ [5.4, 6.5] (72) & $+7635$\,/\,$+8768$\,/\,0 & add $-27$ \\
 & without & 58\,/\,15 & 6356 [5416, 6983] & $+5.4$ [4.8, 5.9] (48) & 0\,/\,$+998$\,/\,$-4002$ & remove $+2179$ \\
\end{tabular}
\end{ruledtabular}
\end{table*}

\subsection{Accompanying foregrounds and spatial binning}
\label{sec:range}
\label{sec:finer}

Across the accompanying foreground choices, the number of structured
\fermi\ bubbles fits passing the original diagnostic ranges from
16 to 132, and median amplitudes range from 2250 to 2475 counts
(Tables~\ref{tab:controls} and \ref{tab:family}). Independent
north/south \fermi\ bubbles normalizations improve the fit by
40--718 and change NFW amplitudes by at most 3.8\% (225 counts) with
the flat template and 450 counts with the structured template, leaving
the quadrant outcomes unchanged. Spatial freedom over the larger eROSITA
footprint has a different effect.

\begin{table}[tbp]
\caption{Accompanying-foreground tests with the structured \fermi\ bubbles
template, Loop~I shells, and $10^\circ$ cells. ``Pass'' requires
$\mathrm{TS}_{20.8}>4$ and four positive quadrant amplitudes. The
last columns give the significance-only count and median 20.8\,GeV
NFW counts. eROSITA denotes the indicated geometric or uniform
bubbles template.}
\label{tab:controls}
\centering
\fontsize{8}{9.5}\selectfont
\setlength{\tabcolsep}{3pt}
\begin{ruledtabular}
\begin{tabular}{lccc}
companions & pass & $>2\sigma$ & counts \\
\hline
solar IC & 16/257 & 183 & 2250 \\
geometric eROSITA outer N/S & 75/257 & 209 & 2475 \\
uniform eROSITA outer N/S & 75/257 & 209 & 2475 \\
uniform eROSITA whole N/S & 132/257 & 181 & 2418 \\
\end{tabular}
\end{ruledtabular}
\end{table}
\begin{table}[tbp]
\caption{Original diagnostic evaluated over twelve sets of accompanying
foreground templates for each \fermi\ bubbles description. The
second column spans the numbers of configurations passing the
combined diagnostic, the third gives the number passing all twelve
sets, and the last spans the numbers above $2\sigma$ without requiring
positive amplitudes in all quadrants. The ensemble contains 257
configurations.}
\label{tab:family}
\centering
\fontsize{8.5}{10.0}\selectfont
\setlength{\tabcolsep}{3pt}
\begin{ruledtabular}
\begin{tabular}{lccc}
bubbles & pass & all 12 & $>2\sigma$ \\
\hline
residual & 155--236 & 155 & 257 \\
flat 2014 & 177--257 & 177 & 211--257 \\
structured 2017 & 0--132 & 0 & 95--221 \\
\end{tabular}
\end{ruledtabular}
\end{table}

The twelve combinations vary Loop~I (shells alone or with restricted
Haslam emission), eROSITA bubbles (absent, uniform, or
brightness-weighted), and solar inverse Compton emission (absent
or present). Solar emission changes the residual-template passing
count from 236 to 155. Brightness-weighted eROSITA templates can
reduce the NFW amplitude fitted with the flat \fermi\ bubbles
template to about 4500 counts.
Individual rather than shared foreground selections change the
residual-template passing count from 236 to 191.

At 20.8\,GeV, independent quadrants improve the coarse fits with residual
\fermi\ bubbles templates by 0.1--1.0 (3.9 for A2020), compared with
12--19 for flat or structured templates. The latter give $\chi^2_3$
tail probabilities of $7.1\times10^{-3}$ to $2.9\times10^{-4}$.
Rotated-partition improvements are 1.8--4.6 and 6--24, respectively;
nine-bin sums are 74--136 and 300--1700. The main text gives the
final calibrated spatial test.

Table~\ref{tab:finer} tests smaller cells before the final PSF
treatment. The NFW amplitudes with the structured \fermi\ bubbles
template increase. These preliminary spectra and likelihood sums
do not replace the final results in Table~\ref{tab:res}.

\begin{table*}[t]
\caption{Spatial-binning tests before PSF folding. The 20.8\,GeV amplitudes
have 68\% profile intervals; TS gives that bin and the nine-bin
sum. Passing the original quadrant diagnostic is shown separately.
The final column compares flat and structured \fermi\ bubbles
templates with NFW allowed and identical accompanying foregrounds.
All likelihood differences use the same binning. Residual templates
are constructed with $10^\circ$ cells. The Galp21 row uses the
isotropic inverse Compton map. These preliminary low-energy
fits are not used for the final inference.}
\label{tab:finer}
\centering
\fontsize{8.3}{9.8}\selectfont
\setlength{\tabcolsep}{3pt}
\begin{ruledtabular}
\begin{tabular}{llccccc}
configuration & set & cells & TS 20.8 / sum & counts [68\%] & pass & flat $-$ struct. \\
\hline
\code{SLZ6R30T150C2} 1-gas & structured & $10^\circ$ & 3.2\,/\,759 & 1181 [516, 1832] & no & $-1114$ \\
 & & $2^\circ$ & 37\,/\,366 & 3262 [2730, 3824] & yes & $-1933$ \\
 & & $1^\circ$ & 46\,/\,376 & 3600 [3045, 4114] & yes & $-1718$ \\
 & flat & $10^\circ$ & 69\,/\,414 & 4668 [4080, 5209] & yes & \\
 & & $1^\circ$ & 118\,/\,584 & 5287 [4786, 5774] & yes & \\
 & residual & $10^\circ$ & 135\,/\,781 & 6074 [5567, 6634] & yes & \\
 & & $1^\circ$ & 287\,/\,1465 & 7762 [7265, 8205] & yes & \\
 & no-residual sel. & $10^\circ$ & 1.3\,/\,957 & 787 [106, 1487] & no & \\
 & & $1^\circ$ & 33\,/\,673 & 3093 [2531, 3614] & yes & \\
\code{SOZ6R30T150C5} 1-gas & structured & $10^\circ$ & 27\,/\,158 & 3037 [2457, 3627] & yes & $-960$ \\
 & & $1^\circ$ & 114\,/\,1107 & 5062 [4552, 5518] & yes & $-1171$ \\
 & residual & $1^\circ$ & 353\,/\,5119 & 7987 [7564, 8455] & yes & \\
 & no-residual sel. & $10^\circ$ & 13\,/\,68 & 2306 [1674, 2973] & no & \\
 & & $1^\circ$ & 68\,/\,279 & 4050 [3570, 4579] & yes & \\
\code{SLZ6R30T150C2} 5-ring & structured & $10^\circ$ & 2.6\,/\,11 & 2700 [1105, 3627] & no & $-1031$ \\
 & & $1^\circ$ & 31\,/\,165 & 5006 [4262, 5747] & yes & $-1470$ \\
 & no-residual sel.\ (flat) & $10^\circ$ & 15\,/\,67 & 6243 [5037, 6937] & yes & \\
 & & $1^\circ$ & 33\,/\,152 & 6356 [5643, 6852] & no & \\
\code{SSZ10R30T150C5} 5-ring & structured & $10^\circ$ & 2.5\,/\,190 & 1575 [587, 2797] & no & $-895$ \\
 & & $1^\circ$ & 23\,/\,76 & 4724 [3784, 5470] & yes & $-1240$ \\
A2020 & structured & $10^\circ$ & 19\,/\,528 & 4556 [3599, 5303] & no & $-711$ \\
 & & $1^\circ$ & 90\,/\,1583 & 6468 [5965, 7016] & yes & $-844$ \\
 & no-residual sel. & $10^\circ$ & 10\,/\,124 & 3487 [2457, 4608] & no & \\
 & & $1^\circ$ & 64\,/\,698 & 5512 [4988, 6079] & yes & \\
Galp21 SA0\_R12 1-gas & structured & $10^\circ$ & 5.2\,/\,393 & 1519 [864, 2206] & no & $-1010$ \\
 & & $1^\circ$ & 54\,/\,222 & 3937 [3412, 4500] & yes & $-1704$ \\
 & residual & $1^\circ$ & 339\,/\,2611 & 8437 [7969, 8904] & yes & \\
\end{tabular}
\end{ruledtabular}
\end{table*}

\section{Simulation and numerical validation}
\label{app:simulations}

\subsection{Amplitude profiling and injection recovery}
\label{app:estimator}

At each trial NFW amplitude, we optimize
the foreground with a convex solver and
refine the amplitude grid near the maximum
and the profile-interval crossings. Three
refinements reduce the grid spacing by a
factor of 64, from 56.2 counts to
approximately 0.88 counts at 20.8\,GeV.
A further refinement leaves the quoted
recovery slopes unchanged. The final
fixed-template controls recover slopes
of 0.99--1.01. Expected counts remain positive,
and all nested likelihood improvements
are nonnegative.

Table~\ref{tab:injection} gives the refined recovery slopes for the
two injected spectra. A constant template coefficient contributes
470 counts at 1.5\,GeV and 30,500 at 100\,GeV for the unit injection;
it is not a flat energy-flux spectrum. Fixed-template fits recover
0.99--1.01, while rebuilding Totani's residual \fermi\ bubbles
templates changes the recovered spectrum. The empirical-spectrum
recovery of 0.93 refers specifically to 20.8\,GeV.

\begin{table}[tbp]
\caption{Recovered-versus-injected slopes after numerical refinement. ``Constant''
denotes a constant template coefficient, not constant $E^2dN/dE$.
``Totani'' denotes the empirical injected spectrum. The residual
\fermi\ bubbles templates are either held fixed or rebuilt after
injection. Dashes indicate bins with no injected signal.}
\label{tab:injection}
\centering\small
\begin{ruledtabular}
\begin{tabular}{rcccc}
& \multicolumn{2}{c}{constant coefficient} & \multicolumn{2}{c}{Totani spectrum}\\
$E$ [GeV] & rebuilt & fixed & rebuilt & fixed\\
\hline
1.5  & $-4.7$ & 1.00 & -- & --\\
2.6  & $-0.8$ & 1.00 & -- & --\\
4.3  & 0.42 & 1.00 & 0.42 & 1.00\\
7.3  & 0.84 & 1.01 & 0.83 & 1.01\\
12.3 & 0.96 & 1.01 & 0.91 & 1.01\\
20.8 & 1.00 & 1.01 & 0.93 & 1.01\\
35   & 0.99 & 0.99 & 0.92 & 0.99\\
59   & 1.00 & 1.00 & 0.92 & 1.01\\
100  & 1.00 & 1.00 & 0.87 & 1.00\\
\end{tabular}
\end{ruledtabular}
\end{table}

\subsection{Null skies and selection experiments}
\label{app:null}

The original selection experiment uses fifty skies per structured
or flat \fermi\ bubbles foreground, with and without the empirical
NFW signal. Each sky is fitted by the 33-configuration subset,
excluding the generating diffuse models and repeating selection.
Twenty-sky controls use the generating diffuse model and substitute
individual templates (Table~\ref{tab:nullsky}). Coarse null spectra
are approximately flat in $E^2dN/dE$, with slopes of $-0.11$ and $-0.09$, compared
with $-0.03$ for the flat-template data spectrum.

\begin{table}[tbp]
\caption{Twenty structured \fermi\ bubbles skies without an NFW signal,
fitted with the generating diffuse model and $10^\circ$ cells.
Counts are medians with 16--84\% simulation ranges. ``Own'' uses
the generating foreground; other rows substitute residual or flat
\fermi\ bubbles templates or use the shared residual-template
selection (``consensus''). Passing refers to the coarse diagnostic.
$\sv$ is in $10^{-25}\,\mathrm{cm^3\,s^{-1}}$ for $b\bar b$ at
500\,GeV.}
\label{tab:nullsky}
\centering
\fontsize{8}{9.5}\selectfont
\setlength{\tabcolsep}{3pt}
\begin{ruledtabular}
\begin{tabular}{lcccc}
set & counts & TS & pass & $\sv$ \\
\hline
own & $-28$ [$-1060$, 551] & 0.5 & 0.00 & $-0.01$ \\
construction & 3977 [3447, 4792] & 55 & 0.60 & $+6.2$ \\
consensus & 3485 [2907, 4445] & 31 & 0.55 & $+5.1$ \\
flat mask & 5109 [4575, 5696] & 97 & 1.00 & $+7.9$ \\
\end{tabular}
\end{ruledtabular}
\end{table}

Replacing structured \fermi\ bubbles with residual templates gives
3977 spurious counts and a 500\,GeV $b\bar b$ coefficient of
$6.2\times10^{-25}\,\mathrm{cm^3\,s^{-1}}$. The flat template gives
5109 counts and $7.9\times10^{-25}$. Both spectra are approximately
flat in $E^2dN/dE$, unlike the full observed spectrum. On flat
\fermi\ bubbles null skies with a five-ring foreground, the
structured template gives 2228 counts at $\mathrm{TS}=2.6$ and a
coefficient near zero, while residual templates give 4927 counts
at $\mathrm{TS}=5.4$. Both generating and fitted morphology matter.

With selection repeated, structured \fermi\ bubbles simulations
recover 5900 and 5600 of 6412 injected counts in single-gas and
five-ring fits; allowing residual templates gives about 9600.
Single-gas fits to a ring-structured null sky return 4700--5600
counts and pass the diagnostic in 94\% of configurations, compared
with 2250 counts and 9\% for five-ring fits. Without residual
templates, all 200 skies select their generating flat or structured
\fermi\ bubbles template. The real-data preferences do not uniquely
identify the morphology.

\subsection{Simulations with finer spatial bins}
\label{app:finer}

The initial ten-sky $2^\circ$ structured \fermi\ bubbles null test
returns $-141\,[-456,425]$ counts with matching templates,
$4577\,[4359,5386]$ with flat templates, and
$4439\,[4378,5117]$ with residual templates (TS 102 and 108 for
the latter two). A matched injection recovers
$6624\,[6085,7191]$ counts for 6412 input, while either mismatch
gives about 11,000. These tests precede the final PSF treatment.

In the twenty-sky PSF-folded selection test, all 33 configurations
select the generating \fermi\ bubbles morphology. Matched-null
medians are $-340$ and $-620$ counts, and injected medians are
5750 and 5440. The separate fifty-sky fixed-configuration calibration
returns 6590 and 6540. These different procedures are not pooled
into one recovery measurement.

The final matched-null $Q^+$ median and 84th percentile over nine
bins are 3--4 and 7--8, consistent with 3.8 and 7.6 from
Eq.~\ref{eq:mixture}. Fitting structured \fermi\ bubbles as flat
instead gives $Q^+=1750\,[1655,1836]$ and about 4700 spurious
counts. Residual templates give $Q^+=870$--1440 and 3300--4300
counts. Matched-null calibration does not account for these
foreground biases.

Fixed-partition null distributions are close to $\chi^2_3$ and
$\chi^2_{27}$ for one and nine bins. The separately calibrated
maximum over partitions has a one-bin median of 3.5--3.7 rather
than 2.4. In coarse controls, a north-only injection gives an
improvement of 83 (72--97), with 2500--3000 northern and 900--1100
southern counts. Matched spherical injections with the structured
\fermi\ bubbles template give a median of 2.1. Residual templates
on structured \fermi\ bubbles skies give medians of 8.4 without
a halo and 8.0 with one. The construction preserves some asymmetry,
but its quadrant response remains foreground-dependent.

\subsection{Noise-free reconstruction}
\label{app:reconstruction}

On a matched, noise-free 4.3\,GeV sky, the 144-cell fit recovers the
input amplitudes to $10^{-6}$. The positive residual image agrees
with the smoothed flat \fermi\ bubbles mask to $6\times10^{-7}$
within the mask. Smoothing moves 6.9\% of its flux beyond the boundary,
and the negative image contains $2\times10^{-7}$ of the positive
flux. This test checks the reconstruction implementation under matched
conditions, separately from the noisy and mismatched simulations.

\section{Extended morphology comparisons}
\label{app:morphology}
\label{app:erosita}

\subsection{eROSITA brightness reconstruction and spatial freedom}

The control footprints have mean longitudes
of $-9.67^\circ$ and $+3.75^\circ$ in the
north and south, respectively, compared
with $-9.46^\circ$ and $+3.75^\circ$ for
the supplied boundaries. We transfer the
reconstructed map using a $960\times960$
crop aligned with the $0.125^\circ$
analysis grid. The East/West brightness
ratios are 0.911 and 0.819 and change by
less than 2\% across the calibration,
registration, and smoothing tests.
Reconstruction from another published
display gives 0.953 and 0.797. Each lobe
is normalized to unit integral over its
support. These checks test the stability
of the reconstruction, but do not constrain
a color-to-intensity error shared by the
published displays.

Individual coarse selections choose brightness-weighted eROSITA
bubbles templates in 128 configurations, no eROSITA component
in 125, and uniform templates in four. Five representative eastern
fits favor brightness weights by 72--319 in $-2\ln\mathcal L$,
while western fits favor uniform emission by 149--326. These
background-only results differ from the final equal-complexity
comparison with NFW included.

\begin{table*}[t]
\caption{eROSITA bubbles tests with flat \fermi\ bubbles, the PSF, and
$2^\circ$ bins. NFW has independent quadrant normalizations.
Counts and south/north ratios are medians over 34 configurations.
West and East denote $\ell<0$ and $\ell>0$. $\Delta$ compares
the nine-bin background-only $-2\ln\mathcal L$ with two uniform
eROSITA lobes; negative is better. The comparison with NFW included
is given in the main text. ``Brightness'' denotes
X-ray-brightness-weighted eROSITA templates.}
\label{tab:erositaflex}
\centering
\fontsize{8.7}{10.2}\selectfont
\setlength{\tabcolsep}{3pt}
\begin{ruledtabular}
\begin{tabular}{lcrcccc}
eROSITA lobes & templates & $\Delta$ & NFW counts & south/north & West & East \\
\hline
none & 0 & $+3061$ & 6079 & 1.12 & 0.94 & 1.35 \\
uniform N/S & 2 & 0 & 5750 & 1.37 & 1.26 & 1.50 \\
brightness N/S & 2 & $+959$ & 5045 & 1.23 & 1.03 & 1.47 \\
uniform N/S$\times$E/W & 4 & $-178$ & 5759 & 1.36 & 1.20 & 1.51 \\
brightness N/S$\times$E/W & 4 & $+732$ & 5232 & 1.35 & 1.08 & 1.60 \\
uniform, 12 pieces & 12 & $-1196$ & 5678 & 1.29 & 1.07 & 1.54 \\
brightness, 12 pieces & 12 & $-1278$ & 4860 & 1.09 & 0.98 & 1.21 \\
uniform, West mirrored & 2 & $+1484$ & 5800 & 1.27 & 1.05 & 1.56 \\
brightness, West mirrored & 2 & $+1792$ & 5176 & 1.11 & 0.88 & 1.40 \\
uniform, West only & 2 & $+1153$ & 5973 & 1.28 & 1.21 & 1.33 \\
brightness, West only & 2 & $+2370$ & 5666 & 1.23 & 1.06 & 1.37 \\
\end{tabular}
\end{ruledtabular}
\end{table*}

The twelve-piece eROSITA bubbles models split the two hemispheres
and east/west regions into $|b|<25^\circ$,
$25^\circ$--$40^\circ$, and $|b|>40^\circ$. With brightness weights,
allowing different hemispheric spectral shapes improves
$-2\ln\mathcal L$ by 5.8 for eight additional parameters
($p=0.67$). Near 21\,GeV, the spectrum reaches approximately
$1.9\times10^{-7}\,\mathrm{GeV\,cm^{-2}\,s^{-1}}$ per hemisphere.
The fitted coefficients are $3.0$ and $4.2\times10^{-25}\,
\mathrm{cm^3\,s^{-1}}$ in the two diffuse decompositions, compared
with $4.9$ and $5.9\times10^{-25}$ for uniform eROSITA bubbles
templates. The low-latitude south-eastern piece contributes 417
counts in Totani's baseline.

Simulated twelve-piece brightness-weighted eROSITA bubbles fitted
with two uniform lobes leave 2500 spurious NFW counts without a
halo. With a spherical halo, the south/north ratio is
$1.35\,[1.29,1.48]$, or 1.34 and 1.41 in West and East.
Matching eROSITA templates recover 6160 of 6412 counts with ratio
1.05, and only 70 without a halo. Foreground mismatch can therefore
produce the asymmetry, without uniquely explaining its origin in
the data.

Simpler null tests vary \fermi\ bubbles morphology, hemispheric
brightness, diffuse model, and accompanying templates. None
reproduces both the observed amplitude and asymmetry. Structured
\fermi\ bubbles fitted as flat give 4200--5100 counts but ratios
of only 1.11--1.20. Omitting eROSITA templates reduces the data
ratio from 1.37 to 1.12. The eastern asymmetry persists without
the reconstructed template. With twelve brightness-weighted pieces,
the eastern ratio is 1.21, compared with $0.97\,[0.80,1.19]$ in
the flexible-template spherical injections.

\subsection{HESTIA projections}
\label{app:hestia}

The HESTIA halos have $c/a=0.64$--0.80 at Galactocentric radii of 1.4--30\,kpc and minor
axes within $5^\circ$ of the disk normal. Observers spaced by
$10^\circ$ around the solar circle give 216 views across the six
halos. We spread each particle's $m\rho/s^2$ contribution over its
32-nearest-neighbor kernel and replace particles within 1\,kpc of
the observer by their mean. The intrinsic south/north ratios relative to NFW are
0.95--1.08, compared with 0.98--1.02 after spherical randomization.
No original view exceeds the NFW ratio by more than 8\%. The fit
comparisons in the main text use these particular halo realizations.

\section{Full-energy-range fits and spectral controls}
\label{app:fullrange}
\label{app:spectralcontrols}

Table~\ref{tab:twodeg} gives all-nine-bin $2^\circ$ selections.
Without residual templates, both decompositions select flat
\fermi\ bubbles, uniform eROSITA bubbles, Haslam, and solar
inverse Compton emission. Removing the three lowest bins yields
the selection in Table~\ref{tab:sel}. Allowing residual templates
selects them in both decompositions and gives larger NFW amplitudes.

\begin{table*}[t]
\caption{Full-nine-bin selections with the PSF and $2^\circ$ cells. Residual
\fermi\ bubbles templates are not convolved again. Shared selection
uses 16 balanced models per decomposition plus A2020; Galp21
models are scored but do not select the shared foregrounds.
Counts are medians and model ranges at 20.8\,GeV. ``Pass'' uses
the original positivity diagnostic. $\sv$ is the median 500\,GeV
$b\bar b$ coefficient in $10^{-25}\,\mathrm{cm^3\,s^{-1}}$.
eROSITA templates are uniform or brightness-weighted.}
\label{tab:twodeg}
\centering
\fontsize{8.5}{10.0}\selectfont
\setlength{\tabcolsep}{3pt}
\begin{ruledtabular}
\begin{tabular}{llcccc}
gas & selected & ensemble & pass & counts [range] & $\sv$ \\
\hline
\multicolumn{6}{l}{\emph{construction available}} \\
single-gas & construction, shells+NPS, eROSITA bright. & 2012 & 17/17 & 6631 [6299, 8704] & $+5.6$ \\
 &  & Galp21 & 12/12 & 7111 [6567, 7535] & $+7.2$ \\
five-ring & construction, shells+NPS, eROSITA unif. & 2012 & 17/17 & 7436 [7022, 8704] & $+7.4$ \\
 &  & Galp21 & 12/12 & 7468 [7015, 7925] & $+7.6$ \\
\multicolumn{6}{l}{\emph{construction excluded}} \\
single-gas & flat mask, eROSITA unif., Haslam, solar IC & 2012 & 17/17 & 5735 [4763, 7404] & $+5.5$ \\
 &  & Galp21 & 12/12 & 5688 [5238, 6100] & $+5.7$ \\
five-ring & flat mask, eROSITA unif., Haslam, solar IC & 2012 & 10/17 & 6205 [5363, 7404] & $+6.0$ \\
 &  & Galp21 & 12/12 & 6695 [6092, 6882] & $+7.2$ \\

\end{tabular}
\end{ruledtabular}
\end{table*}

At $10^\circ$ binning, the full-range 500\,GeV $b\bar b$ coefficients
have medians of $+4.0$ and $+5.9\times10^{-25}\,\mathrm{cm^3\,s^{-1}}$
with residual \fermi\ bubbles templates, and $-2.2$ and
$+5.1\times10^{-25}$ without them (single-gas and five-ring,
respectively). Negative coefficients diagnose the fitted spectral
shape and foreground model; they are not physical cross sections.
With residual templates, the tested $\tau^+\tau^-$ coefficients
have medians of $+6.8$ and $+5.5\times10^{-25}$, and the
$W^+W^-$ coefficients have medians of $+6.0$ and $+8.2\times10^{-25}$.
Without residual templates, the single-gas medians are negative
across these channels ($-1.4$ to $-2.8\times10^{-25}$), while the
five-ring medians are positive ($3.4$--$6.4\times10^{-25}$).
Galp21 coefficients are $+5.7$--$+6.2\times10^{-25}$ under the
imposed residual-template selections and $-1.3$ to
$+1.4\times10^{-25}$ under their individual selections without
residual templates.

The lowest energy bins can dominate these signed spectral fits. In
Totani's baseline without residual \fermi\ bubbles templates,
the first three bins contribute 152, 117, and 80 to the likelihood
improvement for a negative coefficient. In the reported cutoff
control, starting at 5.4\,GeV changes the coefficient from $-4.7$
to $-1.4\times10^{-25}$; the best-fitting single-gas configuration
changes from $+0.3$ to $+1.1\times10^{-25}$. With the residual
templates, the baseline changes from $+3.1$ to $+5.5\times10^{-25}$,
while the five-ring and A2020 controls remain stable within 10\%.
These cutoff controls are separate from
the final six-bin analysis, which repeats
foreground selection and inference over
5.6--130\,GeV with the response included.
For a $b\bar b$ injection matching the
fitted spectral shape, the recovered
normalization changes by less than 2\%
across the tested lower limits of 1.16,
1.9, 3.2, and 5.4\,GeV.

\FloatBarrier
\bibliographystyle{apsrev4-2}
\interlinepenalty=10000
\bibliography{refs}

\begin{thebibliography}{32}%
\makeatletter
\providecommand \@ifxundefined [1]{%
 \@ifx{#1\undefined}
}%
\providecommand \@ifnum [1]{%
 \ifnum #1\expandafter \@firstoftwo
 \else \expandafter \@secondoftwo
 \fi
}%
\providecommand \@ifx [1]{%
 \ifx #1\expandafter \@firstoftwo
 \else \expandafter \@secondoftwo
 \fi
}%
\providecommand \natexlab [1]{#1}%
\providecommand \enquote  [1]{``#1''}%
\providecommand \bibnamefont  [1]{#1}%
\providecommand \bibfnamefont [1]{#1}%
\providecommand \citenamefont [1]{#1}%
\providecommand \href@noop [0]{\@secondoftwo}%
\providecommand \href [0]{\begingroup \@sanitize@url \@href}%
\providecommand \@href[1]{\@@startlink{#1}\@@href}%
\providecommand \@@href[1]{\endgroup#1\@@endlink}%
\providecommand \@sanitize@url [0]{\catcode `\\12\catcode `\$12\catcode
  `\&12\catcode `\#12\catcode `\^12\catcode `\_12\catcode `\%12\relax}%
\providecommand \@@startlink[1]{}%
\providecommand \@@endlink[0]{}%
\providecommand \url  [0]{\begingroup\@sanitize@url \@url }%
\providecommand \@url [1]{\endgroup\@href {#1}{\urlprefix }}%
\providecommand \urlprefix  [0]{URL }%
\providecommand \Eprint [0]{\href }%
\providecommand \doibase [0]{https://doi.org/}%
\providecommand \selectlanguage [0]{\@gobble}%
\providecommand \bibinfo  [0]{\@secondoftwo}%
\providecommand \bibfield  [0]{\@secondoftwo}%
\providecommand \translation [1]{[#1]}%
\providecommand \BibitemOpen [0]{}%
\providecommand \bibitemStop [0]{}%
\providecommand \bibitemNoStop [0]{.\EOS\space}%
\providecommand \EOS [0]{\spacefactor3000\relax}%
\providecommand \BibitemShut  [1]{\csname bibitem#1\endcsname}%
\let\auto@bib@innerbib\@empty
\bibitem [{\citenamefont {Slatyer}(2018)}]{Slatyer:2017sev}%
  \BibitemOpen
  \bibfield  {author} {\bibinfo {author} {\bibfnamefont {T.~R.}\ \bibnamefont
  {Slatyer}},\ }in\ \href {https://doi.org/10.1142/9789813233348_0005} {\emph
  {\bibinfo {booktitle} {{Theoretical Advanced Study Institute in Elementary
  Particle Physics}: {Anticipating the Next Discoveries in Particle
  Physics}}}}\ (\bibinfo {year} {2018})\ pp.\ \bibinfo {pages} {297--353},\
  \Eprint {https://arxiv.org/abs/1710.05137} {arXiv:1710.05137 [hep-ph]}
  \BibitemShut {NoStop}%
\bibitem [{\citenamefont {Ackermann}\ \emph
  {et~al.}(2012{\natexlab{a}})\citenamefont {Ackermann} \emph
  {et~al.}}]{Fermi-LAT:2012pls}%
  \BibitemOpen
  \bibfield  {author} {\bibinfo {author} {\bibfnamefont {M.}~\bibnamefont
  {Ackermann}} \emph {et~al.} (\bibinfo {collaboration} {Fermi-LAT}),\ }\href
  {https://doi.org/10.1088/0004-637X/761/2/91} {\bibfield  {journal} {\bibinfo
  {journal} {Astrophys. J.}\ }\textbf {\bibinfo {volume} {761}},\ \bibinfo
  {pages} {91} (\bibinfo {year} {2012}{\natexlab{a}})},\ \Eprint
  {https://arxiv.org/abs/1205.6474} {arXiv:1205.6474 [astro-ph.CO]}
  \BibitemShut {NoStop}%
\bibitem [{\citenamefont {Chang}\ \emph {et~al.}(2018)\citenamefont {Chang},
  \citenamefont {Lisanti},\ and\ \citenamefont
  {Mishra-Sharma}}]{Chang:2018bpt}%
  \BibitemOpen
  \bibfield  {author} {\bibinfo {author} {\bibfnamefont {L.~J.}\ \bibnamefont
  {Chang}}, \bibinfo {author} {\bibfnamefont {M.}~\bibnamefont {Lisanti}},\
  and\ \bibinfo {author} {\bibfnamefont {S.}~\bibnamefont {Mishra-Sharma}},\
  }\href {https://doi.org/10.1103/PhysRevD.98.123004} {\bibfield  {journal}
  {\bibinfo  {journal} {Phys. Rev. D}\ }\textbf {\bibinfo {volume} {98}},\
  \bibinfo {pages} {123004} (\bibinfo {year} {2018})},\ \Eprint
  {https://arxiv.org/abs/1804.04132} {arXiv:1804.04132 [astro-ph.CO]}
  \BibitemShut {NoStop}%
\bibitem [{\citenamefont {Totani}(2025)}]{Totani:2025fxx}%
  \BibitemOpen
  \bibfield  {author} {\bibinfo {author} {\bibfnamefont {T.}~\bibnamefont
  {Totani}},\ }\href {https://doi.org/10.1088/1475-7516/2025/11/080} {\bibfield
   {journal} {\bibinfo  {journal} {JCAP}\ }\textbf {\bibinfo {volume}
  {2025}}\bibfield  {number} {\bibinfo  {number} { (11)},\ \bibinfo {pages}
  {080}},\ }\Eprint {https://arxiv.org/abs/2507.07209} {arXiv:2507.07209
  [astro-ph.HE]} \BibitemShut {NoStop}%
\bibitem [{\citenamefont {Jho}\ \emph {et~al.}(2025)\citenamefont {Jho},
  \citenamefont {Park}, \citenamefont {Park},\ and\ \citenamefont
  {Park}}]{Jho:2025iah}%
  \BibitemOpen
  \bibfield  {author} {\bibinfo {author} {\bibfnamefont {Y.}~\bibnamefont
  {Jho}}, \bibinfo {author} {\bibfnamefont {J.}~\bibnamefont {Park}}, \bibinfo
  {author} {\bibfnamefont {M.~G.}\ \bibnamefont {Park}},\ and\ \bibinfo
  {author} {\bibfnamefont {S.~C.}\ \bibnamefont {Park}},\ }\href@noop {}
  {\bibinfo {title} {{Aspects of Sommerfeld Enhancement in the light of Halo
  gamma-ray excess}}} (\bibinfo {year} {2025}),\ \Eprint
  {https://arxiv.org/abs/2512.24662} {arXiv:2512.24662 [hep-ph]} \BibitemShut
  {NoStop}%
\bibitem [{\citenamefont {Murayama}(2025)}]{Murayama:2025ihg}%
  \BibitemOpen
  \bibfield  {author} {\bibinfo {author} {\bibfnamefont {H.}~\bibnamefont
  {Murayama}},\ }\href@noop {} {\bibinfo {title} {{Resonant Annihilation of
  WIMP Dark Matter for Halo Gamma Ray Signal}}} (\bibinfo {year} {2025}),\
  \Eprint {https://arxiv.org/abs/2512.01404} {arXiv:2512.01404 [hep-ph]}
  \BibitemShut {NoStop}%
\bibitem [{\citenamefont {Abazajian}\ \emph {et~al.}(2020)\citenamefont
  {Abazajian}, \citenamefont {Horiuchi}, \citenamefont {Kaplinghat},
  \citenamefont {Keeley},\ and\ \citenamefont {Macias}}]{Abazajian:2020tww}%
  \BibitemOpen
  \bibfield  {author} {\bibinfo {author} {\bibfnamefont {K.~N.}\ \bibnamefont
  {Abazajian}}, \bibinfo {author} {\bibfnamefont {S.}~\bibnamefont {Horiuchi}},
  \bibinfo {author} {\bibfnamefont {M.}~\bibnamefont {Kaplinghat}}, \bibinfo
  {author} {\bibfnamefont {R.~E.}\ \bibnamefont {Keeley}},\ and\ \bibinfo
  {author} {\bibfnamefont {O.}~\bibnamefont {Macias}},\ }\href
  {https://doi.org/10.1103/PhysRevD.102.043012} {\bibfield  {journal} {\bibinfo
   {journal} {Phys. Rev. D}\ }\textbf {\bibinfo {volume} {102}},\ \bibinfo
  {pages} {043012} (\bibinfo {year} {2020})},\ \Eprint
  {https://arxiv.org/abs/2003.10416} {arXiv:2003.10416 [hep-ph]} \BibitemShut
  {NoStop}%
\bibitem [{\citenamefont {Abazajian}\ \emph {et~al.}(2026)\citenamefont
  {Abazajian} \emph {et~al.}}]{AbazajianInPrep}%
  \BibitemOpen
  \bibfield  {author} {\bibinfo {author} {\bibfnamefont {K.}~\bibnamefont
  {Abazajian}} \emph {et~al.},\ }\href@noop {} {} (\bibinfo {year} {2026}),\
  \bibinfo {note} {in preparation}\BibitemShut {NoStop}%
\bibitem [{\citenamefont {Cholis}\ \emph {et~al.}(2026)\citenamefont {Cholis},
  \citenamefont {Hu},\ and\ \citenamefont {Zhong}}]{Cholis:2026rba}%
  \BibitemOpen
  \bibfield  {author} {\bibinfo {author} {\bibfnamefont {I.}~\bibnamefont
  {Cholis}}, \bibinfo {author} {\bibfnamefont {L.~Q.}\ \bibnamefont {Hu}},\
  and\ \bibinfo {author} {\bibfnamefont {Y.-M.}\ \bibnamefont {Zhong}},\
  }\href@noop {} {\bibinfo {title} {{The Origin of the Fermi Halo-like
  Emission: A New Model of the Fermi Bubbles}}} (\bibinfo {year} {2026}),\
  \Eprint {https://arxiv.org/abs/2610.10695} {arXiv:2610.10695 [astro-ph.HE]}
  \BibitemShut {NoStop}%
\bibitem [{\citenamefont {Akerib}\ \emph {et~al.}(2026)\citenamefont {Akerib}
  \emph {et~al.}}]{LZ:2026axp}%
  \BibitemOpen
  \bibfield  {author} {\bibinfo {author} {\bibfnamefont {D.~S.}\ \bibnamefont
  {Akerib}} \emph {et~al.} (\bibinfo {collaboration} {LZ}),\ }\href@noop {}
  {\bibinfo {title} {{Search for dark matter particle interactions in an
  extended nuclear recoil energy window with the LUX-ZEPLIN (LZ) experiment}}}
  (\bibinfo {year} {2026}),\ \bibinfo {note} {associated data release:
  \href{https://doi.org/10.17182/hepdata.182472.v1}{HEPData}},\ \Eprint
  {https://arxiv.org/abs/2609.02823} {arXiv:2609.02823 [hep-ex]} \BibitemShut
  {NoStop}%
\bibitem [{\citenamefont {Yamashita}(2026)}]{Yamashita:2026ump}%
  \BibitemOpen
  \bibfield  {author} {\bibinfo {author} {\bibfnamefont {K.}~\bibnamefont
  {Yamashita}},\ }\href@noop {} {\bibinfo {title} {{Inelastic Dark Photon Dark
  Matter for the LUX-ZEPLIN High-Recoil Event and the Galactic Halo Gamma-Ray
  Excess}}} (\bibinfo {year} {2026}),\ \Eprint
  {https://arxiv.org/abs/2609.02868} {arXiv:2609.02868 [hep-ph]} \BibitemShut
  {NoStop}%
\bibitem [{\citenamefont {Su}\ \emph {et~al.}(2010)\citenamefont {Su},
  \citenamefont {Slatyer},\ and\ \citenamefont {Finkbeiner}}]{Su:2010qj}%
  \BibitemOpen
  \bibfield  {author} {\bibinfo {author} {\bibfnamefont {M.}~\bibnamefont
  {Su}}, \bibinfo {author} {\bibfnamefont {T.~R.}\ \bibnamefont {Slatyer}},\
  and\ \bibinfo {author} {\bibfnamefont {D.~P.}\ \bibnamefont {Finkbeiner}},\
  }\href {https://doi.org/10.1088/0004-637X/724/2/1044} {\bibfield  {journal}
  {\bibinfo  {journal} {Astrophys. J.}\ }\textbf {\bibinfo {volume} {724}},\
  \bibinfo {pages} {1044} (\bibinfo {year} {2010})},\ \Eprint
  {https://arxiv.org/abs/1005.5480} {arXiv:1005.5480 [astro-ph.HE]}
  \BibitemShut {NoStop}%
\bibitem [{\citenamefont {Ackermann}\ \emph {et~al.}(2014)\citenamefont
  {Ackermann} \emph {et~al.}}]{Fermi-LAT:2014sfa}%
  \BibitemOpen
  \bibfield  {author} {\bibinfo {author} {\bibfnamefont {M.}~\bibnamefont
  {Ackermann}} \emph {et~al.} (\bibinfo {collaboration} {Fermi-LAT}),\ }\href
  {https://doi.org/10.1088/0004-637X/793/1/64} {\bibfield  {journal} {\bibinfo
  {journal} {Astrophys. J.}\ }\textbf {\bibinfo {volume} {793}},\ \bibinfo
  {pages} {64} (\bibinfo {year} {2014})},\ \Eprint
  {https://arxiv.org/abs/1407.7905} {arXiv:1407.7905 [astro-ph.HE]}
  \BibitemShut {NoStop}%
\bibitem [{\citenamefont {Ackermann}\ \emph {et~al.}(2017)\citenamefont
  {Ackermann} \emph {et~al.}}]{Fermi-LAT:2017opo}%
  \BibitemOpen
  \bibfield  {author} {\bibinfo {author} {\bibfnamefont {M.}~\bibnamefont
  {Ackermann}} \emph {et~al.} (\bibinfo {collaboration} {Fermi-LAT}),\ }\href
  {https://doi.org/10.3847/1538-4357/aa6cab} {\bibfield  {journal} {\bibinfo
  {journal} {Astrophys. J.}\ }\textbf {\bibinfo {volume} {840}},\ \bibinfo
  {pages} {43} (\bibinfo {year} {2017})},\ \Eprint
  {https://arxiv.org/abs/1704.03910} {arXiv:1704.03910 [astro-ph.HE]}
  \BibitemShut {NoStop}%
\bibitem [{\citenamefont {Wolleben}(2007)}]{Wolleben:2007pq}%
  \BibitemOpen
  \bibfield  {author} {\bibinfo {author} {\bibfnamefont {M.}~\bibnamefont
  {Wolleben}},\ }\href {https://doi.org/10.1086/518711} {\bibfield  {journal}
  {\bibinfo  {journal} {Astrophys. J.}\ }\textbf {\bibinfo {volume} {664}},\
  \bibinfo {pages} {349} (\bibinfo {year} {2007})},\ \Eprint
  {https://arxiv.org/abs/0704.0276} {arXiv:0704.0276 [astro-ph]} \BibitemShut
  {NoStop}%
\bibitem [{\citenamefont {Predehl}\ \emph {et~al.}(2020)\citenamefont {Predehl}
  \emph {et~al.}}]{Predehl:2020kyq}%
  \BibitemOpen
  \bibfield  {author} {\bibinfo {author} {\bibfnamefont {P.}~\bibnamefont
  {Predehl}} \emph {et~al.},\ }\href
  {https://doi.org/10.1038/s41586-020-2979-0} {\bibfield  {journal} {\bibinfo
  {journal} {Nature}\ }\textbf {\bibinfo {volume} {588}},\ \bibinfo {pages}
  {227} (\bibinfo {year} {2020})},\ \Eprint {https://arxiv.org/abs/2012.05840}
  {arXiv:2012.05840 [astro-ph.GA]} \BibitemShut {NoStop}%
\bibitem [{\citenamefont {Foster}(2026)}]{Foster:2026cern}%
  \BibitemOpen
  \bibfield  {author} {\bibinfo {author} {\bibfnamefont {J.~W.}\ \bibnamefont
  {Foster}},\ }\href@noop {} {\bibinfo {title} {{The 20 GeV Gamma-Ray Excess is
  not from Dark Matter}}},\ \bibinfo {howpublished} {Talk at CERN, 21 July
  2026} (\bibinfo {year} {2026})\BibitemShut {NoStop}%
\bibitem [{\citenamefont {Atwood}\ \emph {et~al.}(2009)\citenamefont {Atwood}
  \emph {et~al.}}]{Atwood:2009ez}%
  \BibitemOpen
  \bibfield  {author} {\bibinfo {author} {\bibfnamefont {W.~B.}\ \bibnamefont
  {Atwood}} \emph {et~al.} (\bibinfo {collaboration} {Fermi-LAT}),\ }\href
  {https://doi.org/10.1088/0004-637X/697/2/1071} {\bibfield  {journal}
  {\bibinfo  {journal} {Astrophys. J.}\ }\textbf {\bibinfo {volume} {697}},\
  \bibinfo {pages} {1071} (\bibinfo {year} {2009})},\ \Eprint
  {https://arxiv.org/abs/0902.1089} {arXiv:0902.1089 [astro-ph.IM]}
  \BibitemShut {NoStop}%
\bibitem [{\citenamefont {Ackermann}\ \emph
  {et~al.}(2012{\natexlab{b}})\citenamefont {Ackermann} \emph
  {et~al.}}]{Ackermann:2012pya}%
  \BibitemOpen
  \bibfield  {author} {\bibinfo {author} {\bibfnamefont {M.}~\bibnamefont
  {Ackermann}} \emph {et~al.} (\bibinfo {collaboration} {Fermi-LAT}),\ }\href
  {https://doi.org/10.1088/0004-637X/750/1/3} {\bibfield  {journal} {\bibinfo
  {journal} {Astrophys. J.}\ }\textbf {\bibinfo {volume} {750}},\ \bibinfo
  {pages} {3} (\bibinfo {year} {2012}{\natexlab{b}})},\ \Eprint
  {https://arxiv.org/abs/1202.4039} {arXiv:1202.4039 [astro-ph.HE]}
  \BibitemShut {NoStop}%
\bibitem [{\citenamefont {Porter}\ \emph {et~al.}(2022)\citenamefont {Porter},
  \citenamefont {Johannesson},\ and\ \citenamefont
  {Moskalenko}}]{Porter:2021tlr}%
  \BibitemOpen
  \bibfield  {author} {\bibinfo {author} {\bibfnamefont {T.~A.}\ \bibnamefont
  {Porter}}, \bibinfo {author} {\bibfnamefont {G.}~\bibnamefont
  {Johannesson}},\ and\ \bibinfo {author} {\bibfnamefont {V.~I.}\ \bibnamefont
  {Moskalenko}},\ }\href {https://doi.org/10.3847/1538-4365/ac80f6} {\bibfield
  {journal} {\bibinfo  {journal} {Astrophys. J. Supp.}\ }\textbf {\bibinfo
  {volume} {262}},\ \bibinfo {pages} {30} (\bibinfo {year} {2022})},\ \Eprint
  {https://arxiv.org/abs/2112.12745} {arXiv:2112.12745 [astro-ph.HE]}
  \BibitemShut {NoStop}%
\bibitem [{\citenamefont {Porter}\ \emph {et~al.}(2017)\citenamefont {Porter},
  \citenamefont {Johannesson},\ and\ \citenamefont
  {Moskalenko}}]{Porter:2017vaa}%
  \BibitemOpen
  \bibfield  {author} {\bibinfo {author} {\bibfnamefont {T.~A.}\ \bibnamefont
  {Porter}}, \bibinfo {author} {\bibfnamefont {G.}~\bibnamefont
  {Johannesson}},\ and\ \bibinfo {author} {\bibfnamefont {I.~V.}\ \bibnamefont
  {Moskalenko}},\ }\href {https://doi.org/10.3847/1538-4357/aa844d} {\bibfield
  {journal} {\bibinfo  {journal} {Astrophys. J.}\ }\textbf {\bibinfo {volume}
  {846}},\ \bibinfo {pages} {67} (\bibinfo {year} {2017})},\ \Eprint
  {https://arxiv.org/abs/1708.00816} {arXiv:1708.00816 [astro-ph.HE]}
  \BibitemShut {NoStop}%
\bibitem [{\citenamefont {Abdollahi}\ \emph {et~al.}(2020)\citenamefont
  {Abdollahi} \emph {et~al.}}]{Fermi-LAT:2019yla}%
  \BibitemOpen
  \bibfield  {author} {\bibinfo {author} {\bibfnamefont {S.}~\bibnamefont
  {Abdollahi}} \emph {et~al.} (\bibinfo {collaboration} {Fermi-LAT}),\ }\href
  {https://doi.org/10.3847/1538-4365/ab6bcb} {\bibfield  {journal} {\bibinfo
  {journal} {Astrophys. J. Suppl.}\ }\textbf {\bibinfo {volume} {247}},\
  \bibinfo {pages} {33} (\bibinfo {year} {2020})},\ \Eprint
  {https://arxiv.org/abs/1902.10045} {arXiv:1902.10045 [astro-ph.HE]}
  \BibitemShut {NoStop}%
\bibitem [{\citenamefont {Ballet}\ \emph {et~al.}(2023)\citenamefont {Ballet},
  \citenamefont {Bruel}, \citenamefont {Burnett},\ and\ \citenamefont
  {Lott}}]{Ballet:2023qzs}%
  \BibitemOpen
  \bibfield  {author} {\bibinfo {author} {\bibfnamefont {J.}~\bibnamefont
  {Ballet}}, \bibinfo {author} {\bibfnamefont {P.}~\bibnamefont {Bruel}},
  \bibinfo {author} {\bibfnamefont {T.~H.}\ \bibnamefont {Burnett}},\ and\
  \bibinfo {author} {\bibfnamefont {B.}~\bibnamefont {Lott}} (\bibinfo
  {collaboration} {Fermi-LAT}),\ }\href@noop {} {\bibinfo {title} {{Fermi Large
  Area Telescope Fourth Source Catalog Data Release 4 (4FGL-DR4)}}} (\bibinfo
  {year} {2023}),\ \Eprint {https://arxiv.org/abs/2307.12546} {arXiv:2307.12546
  [astro-ph.HE]} \BibitemShut {NoStop}%
\bibitem [{\citenamefont {Remazeilles}\ \emph {et~al.}(2015)\citenamefont
  {Remazeilles}, \citenamefont {Dickinson}, \citenamefont {Banday},
  \citenamefont {Bigot-Sazy},\ and\ \citenamefont
  {Ghosh}}]{Remazeilles:2014mba}%
  \BibitemOpen
  \bibfield  {author} {\bibinfo {author} {\bibfnamefont {M.}~\bibnamefont
  {Remazeilles}}, \bibinfo {author} {\bibfnamefont {C.}~\bibnamefont
  {Dickinson}}, \bibinfo {author} {\bibfnamefont {A.~J.}\ \bibnamefont
  {Banday}}, \bibinfo {author} {\bibfnamefont {M.~A.}\ \bibnamefont
  {Bigot-Sazy}},\ and\ \bibinfo {author} {\bibfnamefont {T.}~\bibnamefont
  {Ghosh}},\ }\href {https://doi.org/10.1093/mnras/stv1274} {\bibfield
  {journal} {\bibinfo  {journal} {Mon. Not. Roy. Astron. Soc.}\ }\textbf
  {\bibinfo {volume} {451}},\ \bibinfo {pages} {4311} (\bibinfo {year}
  {2015})},\ \Eprint {https://arxiv.org/abs/1411.3628} {arXiv:1411.3628
  [astro-ph.IM]} \BibitemShut {NoStop}%
\bibitem [{\citenamefont {Zhang}\ \emph {et~al.}(2024)\citenamefont {Zhang}
  \emph {et~al.}}]{Zhang:2024mcd}%
  \BibitemOpen
  \bibfield  {author} {\bibinfo {author} {\bibfnamefont {H.-S.}\ \bibnamefont
  {Zhang}} \emph {et~al.},\ }\href {https://doi.org/10.1038/s41550-024-02362-0}
  {\bibfield  {journal} {\bibinfo  {journal} {Nature Astron.}\ }\textbf
  {\bibinfo {volume} {8}},\ \bibinfo {pages} {1416} (\bibinfo {year} {2024})},\
  \Eprint {https://arxiv.org/abs/2408.06312} {arXiv:2408.06312 [astro-ph.GA]}
  \BibitemShut {NoStop}%
\bibitem [{\citenamefont {Macias}\ \emph {et~al.}(2018)\citenamefont {Macias},
  \citenamefont {Gordon}, \citenamefont {Crocker}, \citenamefont {Coleman},
  \citenamefont {Paterson}, \citenamefont {Horiuchi},\ and\ \citenamefont
  {Pohl}}]{Macias:2016nev}%
  \BibitemOpen
  \bibfield  {author} {\bibinfo {author} {\bibfnamefont {O.}~\bibnamefont
  {Macias}}, \bibinfo {author} {\bibfnamefont {C.}~\bibnamefont {Gordon}},
  \bibinfo {author} {\bibfnamefont {R.~M.}\ \bibnamefont {Crocker}}, \bibinfo
  {author} {\bibfnamefont {B.}~\bibnamefont {Coleman}}, \bibinfo {author}
  {\bibfnamefont {D.}~\bibnamefont {Paterson}}, \bibinfo {author}
  {\bibfnamefont {S.}~\bibnamefont {Horiuchi}},\ and\ \bibinfo {author}
  {\bibfnamefont {M.}~\bibnamefont {Pohl}},\ }\href
  {https://doi.org/10.1038/s41550-018-0414-3} {\bibfield  {journal} {\bibinfo
  {journal} {Nature Astron.}\ }\textbf {\bibinfo {volume} {2}},\ \bibinfo
  {pages} {387} (\bibinfo {year} {2018})},\ \Eprint
  {https://arxiv.org/abs/1611.06644} {arXiv:1611.06644 [astro-ph.HE]}
  \BibitemShut {NoStop}%
\bibitem [{\citenamefont {Cowan}\ \emph {et~al.}(2011)\citenamefont {Cowan},
  \citenamefont {Cranmer}, \citenamefont {Gross},\ and\ \citenamefont
  {Vitells}}]{Cowan:2010js}%
  \BibitemOpen
  \bibfield  {author} {\bibinfo {author} {\bibfnamefont {G.}~\bibnamefont
  {Cowan}}, \bibinfo {author} {\bibfnamefont {K.}~\bibnamefont {Cranmer}},
  \bibinfo {author} {\bibfnamefont {E.}~\bibnamefont {Gross}},\ and\ \bibinfo
  {author} {\bibfnamefont {O.}~\bibnamefont {Vitells}},\ }\href@noop {}
  {\bibfield  {journal} {\bibinfo  {journal} {Eur. Phys. J. C}\ }\textbf
  {\bibinfo {volume} {71}},\ \bibinfo {pages} {1554} (\bibinfo {year}
  {2011})},\ \Eprint {https://arxiv.org/abs/1007.1727} {arXiv:1007.1727
  [physics.data-an]} \BibitemShut {NoStop}%
\bibitem [{\citenamefont {Schwarz}(1978)}]{Schwarz:1978bic}%
  \BibitemOpen
  \bibfield  {author} {\bibinfo {author} {\bibfnamefont {G.}~\bibnamefont
  {Schwarz}},\ }\href {https://doi.org/10.1214/aos/1176344136} {\bibfield
  {journal} {\bibinfo  {journal} {Ann. Statist.}\ }\textbf {\bibinfo {volume}
  {6}},\ \bibinfo {pages} {461} (\bibinfo {year} {1978})}\BibitemShut {NoStop}%
\bibitem [{\citenamefont {Cirelli}\ \emph {et~al.}(2011)\citenamefont
  {Cirelli}, \citenamefont {Corcella}, \citenamefont {Hektor}, \citenamefont
  {Hutsi}, \citenamefont {Kadastik}, \citenamefont {Panci}, \citenamefont
  {Raidal}, \citenamefont {Sala},\ and\ \citenamefont
  {Strumia}}]{Cirelli:2010xx}%
  \BibitemOpen
  \bibfield  {author} {\bibinfo {author} {\bibfnamefont {M.}~\bibnamefont
  {Cirelli}}, \bibinfo {author} {\bibfnamefont {G.}~\bibnamefont {Corcella}},
  \bibinfo {author} {\bibfnamefont {A.}~\bibnamefont {Hektor}}, \bibinfo
  {author} {\bibfnamefont {G.}~\bibnamefont {Hutsi}}, \bibinfo {author}
  {\bibfnamefont {M.}~\bibnamefont {Kadastik}}, \bibinfo {author}
  {\bibfnamefont {P.}~\bibnamefont {Panci}}, \bibinfo {author} {\bibfnamefont
  {M.}~\bibnamefont {Raidal}}, \bibinfo {author} {\bibfnamefont
  {F.}~\bibnamefont {Sala}},\ and\ \bibinfo {author} {\bibfnamefont
  {A.}~\bibnamefont {Strumia}},\ }\href
  {https://doi.org/10.1088/1475-7516/2011/03/051} {\bibfield  {journal}
  {\bibinfo  {journal} {JCAP}\ }\textbf {\bibinfo {volume} {2011}}\bibfield
  {number} {\bibinfo  {number} { (03)},\ \bibinfo {pages} {051}},\ }\bibinfo
  {note} {[Erratum: JCAP 10, E01 (2012)]},\ \Eprint
  {https://arxiv.org/abs/1012.4515} {arXiv:1012.4515 [hep-ph]} \BibitemShut
  {NoStop}%
\bibitem [{\citenamefont {{Libeskind}}\ \emph {et~al.}(2020)\citenamefont
  {{Libeskind}} \emph {et~al.}}]{2020MNRAS.498.2968L}%
  \BibitemOpen
  \bibfield  {author} {\bibinfo {author} {\bibfnamefont {N.~I.}\ \bibnamefont
  {{Libeskind}}} \emph {et~al.},\ }\href
  {https://doi.org/10.1093/mnras/staa2541} {\bibfield  {journal} {\bibinfo
  {journal} {Mon. Not. Roy. Astron. Soc.}\ }\textbf {\bibinfo {volume} {498}},\
  \bibinfo {pages} {2968} (\bibinfo {year} {2020})},\ \Eprint
  {https://arxiv.org/abs/2008.04926} {arXiv:2008.04926 [astro-ph.GA]}
  \BibitemShut {NoStop}%
\bibitem [{\citenamefont {Muru}\ \emph {et~al.}(2025)\citenamefont {Muru},
  \citenamefont {Silk}, \citenamefont {Libeskind}, \citenamefont {Gottloeber},\
  and\ \citenamefont {Hoffman}}]{Muru:2025vpz}%
  \BibitemOpen
  \bibfield  {author} {\bibinfo {author} {\bibfnamefont {M.~M.}\ \bibnamefont
  {Muru}}, \bibinfo {author} {\bibfnamefont {J.}~\bibnamefont {Silk}}, \bibinfo
  {author} {\bibfnamefont {N.~I.}\ \bibnamefont {Libeskind}}, \bibinfo {author}
  {\bibfnamefont {S.}~\bibnamefont {Gottloeber}},\ and\ \bibinfo {author}
  {\bibfnamefont {Y.}~\bibnamefont {Hoffman}},\ }\href
  {https://doi.org/10.1103/g9qz-h8wd} {\bibfield  {journal} {\bibinfo
  {journal} {Phys. Rev. Lett.}\ }\textbf {\bibinfo {volume} {135}},\ \bibinfo
  {pages} {161005} (\bibinfo {year} {2025})},\ \Eprint
  {https://arxiv.org/abs/2508.06314} {arXiv:2508.06314 [astro-ph.HE]}
  \BibitemShut {NoStop}%
\bibitem [{\citenamefont {J{\'o}hannesson}\ \emph {et~al.}(2018)\citenamefont
  {J{\'o}hannesson}, \citenamefont {Porter},\ and\ \citenamefont
  {Moskalenko}}]{Johannesson:2018bit}%
  \BibitemOpen
  \bibfield  {author} {\bibinfo {author} {\bibfnamefont {G.}~\bibnamefont
  {J{\'o}hannesson}}, \bibinfo {author} {\bibfnamefont {T.~A.}\ \bibnamefont
  {Porter}},\ and\ \bibinfo {author} {\bibfnamefont {I.~V.}\ \bibnamefont
  {Moskalenko}},\ }\href {https://doi.org/10.3847/1538-4357/aab26e} {\bibfield
  {journal} {\bibinfo  {journal} {Astrophys. J.}\ }\textbf {\bibinfo {volume}
  {856}},\ \bibinfo {pages} {45} (\bibinfo {year} {2018})},\ \Eprint
  {https://arxiv.org/abs/1802.08646} {arXiv:1802.08646 [astro-ph.HE]}
  \BibitemShut {NoStop}%
\end{thebibliography}%
\end{document}